\documentclass[conference]{IEEEtran}
\ifCLASSINFOpdf
\else
\fi

\usepackage{amssymb}
\usepackage{amsmath}
\usepackage{mathptmx}
\usepackage{graphicx}
\usepackage{algorithmic}
\usepackage{algorithm}
\usepackage{subfigure}
\usepackage{times}
\usepackage{url}
\usepackage{multirow}

\newtheorem{theorem}{Theorem}

\newcommand{\eat}[1]{}

\begin{document}
\title{DRS: Dynamic Resource Scheduling for Real-Time Analytics over Fast Streams}

%\author{\IEEEauthorblockN{Michael Shell}
%\IEEEauthorblockA{School of Electrical and\\Computer Engineering\\
%Georgia Institute of Technology\\
%Atlanta, Georgia 30332--0250\\
%Email: http://www.michaelshell.org/contact.html}
%\and
%\IEEEauthorblockN{Homer Simpson}
%\IEEEauthorblockA{Twentieth Century Fox\\
%\and
%\IEEEauthorblockN{James Kirk\\ and Montgomery Scott}
%\IEEEauthorblockA{Starfleet Academy\\
%San Francisco, California 96678-2391\\

\author{
   \IEEEauthorblockN{Tom Z. J. Fu$^1$~Jianbing Ding$^2$~Richard T. B. Ma$^{1,3}$~Marianne Winslett$^4$~Yin Yang$^5$~Zhenjie Zhang$^1$}
   \IEEEauthorblockA{$^1$Advanced Digital Sciences Center, Illinois at Singapore Pte. Ltd.\\
   $^2$School of Information Science and Technology, Sun Yat-sen University\\
   $^3$School of Computing, National University of Singapore\\
   $^4$Department of Computer Science, University of Illinois at Urbana-Champaign\\
   $^5$College of Science and Engineering, Hamad Bin Khalifa University\\
   Email: \{tom.fu,zhenjie\}@adsc.com.sg, dingsword@gmail.com, tbma@comp.nus.edu.sg, winslett@illinois.edu, yyang@qf.org.qa}
}


%Telephone: (800) 555--1212, Fax: (888) 555--1212}


\maketitle
%%%%%%%%%%%%%%%%%%%%%%%%%%%%%%%%%%%%%%%%%%%%%%%%%%%%
%%
%%   dvips -P cmz -t a4 -o <file>.ps <file>.dvi
%%   dvips -P cmz -t a4 -o drs_vldb.ps drs_vldb.dvi
%%
%%%%%%%%%%%%%%%%%%%%%%%%%%%%%%%%%%%%%%%%%%%%%%%%%%%%%
\begin{abstract}
In a data stream management system (DSMS), users register continuous queries, and receive result updates as data arrive and expire. We focus on applications with real-time constraints, in which the user must receive each result update within a given period after the update occurs. To handle fast data, the DSMS is commonly placed on top of a cloud infrastructure. Because stream properties such as arrival rates can fluctuate unpredictably, cloud resources must be dynamically provisioned and scheduled accordingly to ensure real-time response. It is essential, for the existing systems or future developments, to possess the ability of scheduling resources dynamically according to the current workload, in order to avoid wasting resources, or failing in delivering correct results on time.

Motivated by this, we propose DRS, a novel dynamic resource scheduler for cloud-based DSMSs. DRS overcomes three fundamental challenges: (a) how to model the relationship between the provisioned resources and query response time (b) where to best place resources; and (c) how to measure system load with minimal overhead. In particular, DRS includes an accurate performance model based on the theory of \emph{Jackson open queueing networks} and is capable of handling \emph{arbitrary} operator topologies, possibly with loops, splits and joins. Extensive experiments with real data confirm that DRS achieves real-time response with close to optimal resource consumption.
\end{abstract}

\IEEEpeerreviewmaketitle

\section{Introduction}
In many applications, such as analytics over microblogs, video feeds and sensor readings, data records are not available beforehand, but gradually and continuously arrive in the form of streams. A data stream management system (DSMS) handles such streams, and answers long-running, continuous queries to users. The results of such a query are delivered in the form of a stream of updates. Often, users are interested in performing streaming analytics in \emph{real time}, meaning that each result update must reach the user within a given time period after the update occurs, i.e., the earliest possible time that it can be produced. For instance, consider a DSMS monitoring surveillance video streams in hospital wards. Events such as a patient falling should be detected promptly to alarm doctors and nurses in time.

To deal with fast, high-volume streams and stringent real-time response requirements, it is increasingly common to place the DSMS on top of a cloud infrastructure, which provides virtually unlimited computing resources on demand. Because key properties of a data stream, including its volume, arrival rates, value distribution, etc., can fluctuate in an unpredictable manner, the DSMS should ideally dynamically provision cloud resources to each application, in order to satisfy the real-time constraints with minimum resource consumption. Meanwhile, inside an application, resources need to be carefully \emph{scheduled} to different components to ensure optimal utilization. Misplacing resources may cause not only poor resource utilization, but instability of the system as a whole.

\begin{figure}[htb]
    \centering
    \vspace{-5pt}
    \includegraphics[width=3.4in]{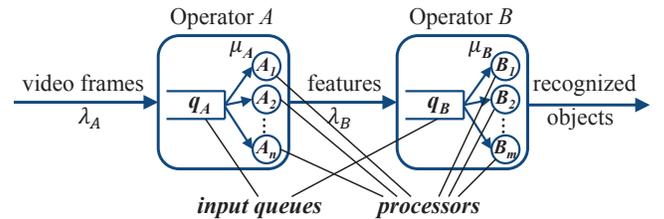}
    \caption{Example streaming analytics application.}
    \vspace{-5pt}
    \label{Fig:runExp1}
\end{figure}

Figure~\ref{Fig:runExp1} shows an example video stream processing application with two operators $A$ (which extracts features from input video frames) and $B$ (which recognizes objects from the extracted features), with the output of $A$ fed to $B$ as input. The record arrival rates for $A$ and $B$ are $\lambda_A$ and $\lambda_B$ respectively, where $\lambda_A$ depends on the input, e.g., 24 frames per second, while $\lambda_B$ depends on the output rate of $A$, i.e., the number of features extracted in unit time. Inside each operator, an input is first buffered into an \emph{input queue} (i.e., $q_A$ in $A$ and $q_B$ in $B$) before being processed by one of the parallel processors ($A_1,\dots, A_n$ in $A$, $B_1,\dots,B_m$ in $B$). Assuming the cloud provides identical processing units, each processor in $A$ (respectively in $B$) can process $\mu_A$ ($\mu_B$) inputs in a unit of time. Clearly, an operator must have sufficient processors to keep up with its input rate; otherwise, inputs start to fill its input queue, leading to increased latency due to waiting, and, eventually, errors when the queue reaches its size limit. Since the data arrival rate and processing rate for each processor are uncontrollable, the main resource scheduling issue is to determine the number of processors in each operator, in our example, $n$ and $m$\footnote{Although there are other types of cloud resources, such as storage and network bandwidth, we focus on computation-intensive applications where processors are the key resource.}.

\begin{figure}[htb]
    \centering
    \vspace{-5pt}
    \includegraphics[width=2.4in]{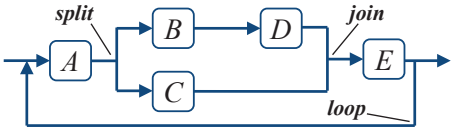}
    \caption{Example complex operator topology.}
    \vspace{-5pt}
    \label{Fig:runExp2}
\end{figure}

A simple approach to scheduling resources is to monitor the workload in each operator, and adjust the number of processors accordingly. This method is insufficient in multi-operator applications. For instance, consider the case that at some point, many recognizable objects appear in the video stream. Then, although the number of frames per second in the input (i.e., $\lambda_A$) remains stable, each frame now contains more extractable features, requiring more work at operator $A$. Hence, $\mu_A$ decreases, which consequently overloads operator $A$, causing inputs to wait longer in its queue $q_A$, slowing down query response. Now, if we naively add processors to $A$ to flush $q_A$, operator $A$ then suddenly produces a large amount of outputs, leading to a burst in the input rate $\lambda_B$ of operator $B$, overloading the latter. This problem is exacerbated when the application involves a complex network of operators. Figure~\ref{Fig:runExp2} shows such an example, with splits ($A$ to $B$, $C$), joins ($C$, $D$ to $E$) and a feedback loop ($E$ to $A$). Such topological features are key enablers for certain applications, e.g., loops allows data reduction at the input based on the current query results, as we show with an example in Section~\ref{Sec:Exp}.

As we review in Section~\ref{sec:related}, existing systems largely overlook the problem of dynamic resource scheduling. Consequently, to meet the real-time constraint, they either require manual tuning at runtime (which is infeasible for dynamic streams), overprovisioning resources to each operator (which wastes resources), or load shedding (which leads to incorrect results).
Motivated by this, we design and implement DRS, a dynamic resource scheduling module. DRS generally applies to operator-based DSMSs,
and allows operators to form an \emph{arbitrary} topology, possibly with splits, joins and loops as shown in Figure~\ref{Fig:runExp2}. In particular, the support for loops can be a key enabler for certain applications, especially those involving iterations, as we show with an example in Section~\ref{Sec:Exp}. Meanwhile, from a semantics point of view, allowing arbitrary topologies is more general than two-step MapUpdate in Muppet~\cite{lam2012muppet}, and the DAG model in TimeStream~\cite{qian2013timestream}.

Our main contributions include effective and efficient solutions to three fundamental problems in dynamic resource scheduling: (a) how much resources are needed, (b) where to best place the allocated resources to minimize response time, and (c) how to implement resource scheduling in a real system with minimal overhead. In particular, our solutions to the first two problems are based on the theory of extended Jackson networks, which provides an educated estimate of system performance.

The rest of the paper is organized as follows. Section~\ref{sec:related} surveys related work. Section~\ref{sec:drs} presents our performance model and optimization algorithm. Section~\ref{Sec:sysDesign} describes the implementation of DRS. Section~\ref{Sec:Exp} contains an extensive set of experiments with real data. Section~\ref{Sec:Conclusion} concludes with directions for future work.
%

%\input{related}
%\newpage
\section{Related Work}\label{sec:related}
%Section~\ref{Sec:related:rsCloud} overviews general resource scheduling in cloud-based systems. Section~\ref{Sec:related:TradDSMS} reviews traditional DSMSs that predates cloud-based ones. Section~\ref{Sec:related:CBSP} describes cloud-based DSMSs.

\subsection{Resource Scheduling in Cloud Systems}\label{Sec:related:rsCloud}
A cloud consists of a massive number of interconnected commodity servers. A key feature of the cloud is that its resources, such as CPU cores, memory, disk space and network bandwidth can be provisioned to applications on demand. In fact, most cloud infrastructure providers today offer pay-as-you-go options for resource usage. Hence, a fundamental requirement for a system to effectively use the cloud is \emph{elasticity}, meaning that the system must be able to dynamically allocate and release cloud resources based on the current workload. Many traditional parallel and distributed systems, however, assume a fixed amount of resources available beforehand, rendering them unsuitable to be applied in a cloud platform. As a result, many novel elastic cloud-based paradigms and systems have emerged in the past decade.

The first wave of cloud-based systems were built for running a batch of (often slow) jobs offline. Notably, MapReduce \cite{dean2008mapreduce} is a batch processing framework that hides the complexity of the cloud infrastructure, and exposes a simple programming interface to users consisting of two functions: \emph{map} (e.g., for data filtering and transformation) and \emph{reduce} (for aggregation and join). A plethora of MapReduce systems, improvements, techniques, and optimizations have been proposed in recent years, and we refer the reader to a comprehensive survey \cite{li2014distributed}.

Resource scheduling has been a central problem in Map-Reduce like systems, and a plethora of schedulers have been developed and used in production, e.g., Fair Scheduler~\cite{fairScheduler}, Capacity Schedular~\cite{capacityScheduler}. Since tasks running on nodes without relevant data incur costly network transmissions, delay scheduling~\cite{zaharia2010delay} reduces such non-local tasks by forcing nodes to wait until either a local task appears, or a specified period has passed. These scheduling strategies, however, do not apply to our problem, because they are designed for offline, batch processing of (semi-) static data, where the goal is to minimize \emph{total} job completion time; in contrast, we focus on real-time processing of streaming data, where each \emph{individual} result update must be delivered on time.

Recently, much attention has been shifted to real-time interactive systems for big data analytics, such as Dremel~\cite{melnik2010dremel},
%Impala~\cite{kornacker2012cloudera},
Impala, Presto~\cite{presto}, OceanRT~\cite{zhang2014oceanrt, Zhang2014oceanrt2}, C-Cube~\cite{zhang2013c}, SADA~\cite{cai2013sada} and newer versions of Hive~\cite{thusoo2010hive}. Such systems deal with static rather than streaming data; meanwhile, the term ``real-time'' here has a different meaning: that each query is executed quickly enough so that the user can wait online for its results. Hence, resource scheduling in these systems resembles offline systems, and their techniques do not apply to our problem for similar reasons. Another recent hot topic in cloud-based system research is cloud-based stream processing, which is most relevant to this work. We review them in Section~\ref{Sec:related:CBSP}.

Finally, there exist generic scheduling solutions for provisioning to multiple applications competing for cloud resources. System such as Mesos~\cite{hindman2011mesos}, YARN~\cite{yarn} are prominent examples. Abacus~\cite{zhang2013abacus} optimizes total utility by allocating resources via a truth-revealing auction. These methods generally assume that an application already knows the amount of resources it needs, and how to distribute these resources internally, which are the problems solved in this paper. Hence, they can be used in combination with the proposed solution.

%\newpage
\subsection{Traditional DSMSs}\label{Sec:related:TradDSMS}
Stream processing has been an important research topic in both academia and industry. Earlier work focuses on DSMSs in a centralized setting, which resembles the traditional, centralized database management systems. For instance, STREAM~\cite{arasu2003stream} establishes formal semantics for queries over streams~\cite{arasu2006cql}, and proposes efficient query processing algorithms, e.g.,~\cite{babu2005adaptive}. Similar systems include Aurora~\cite{abadi2003aurora}, Gigascope~\cite{cranor2003gigascope}, TelegraphCQ~\cite{chandrasekaran2003telegraphcq}, and System S~\cite{andrade2011processing}. Scheduling in such centralized systems means deciding the best order of operators to execute (by the central processor), e.g., in order to minimize memory consumption~\cite{babcock2004operator}. Hence, scheduling strategies in these systems such as~\cite{babcock2004operator} do not apply to our cloud-based setting, where operators are executed by multiple nodes in parallel, and computational resources are dynamically provisioned on demand.

Similarly, DSMSs built for traditional parallel settings, notably Borealis~\cite{abadi2005design}, also differ from cloud-based DSMSs in that the former assume that a fixed amount of computational resources available beforehand, rather than dynamically allocated. Hence, to our knowledge, no scheduling technique along this line of research applies to our problem. Next we review cloud-based DSMSs.

\subsection{Cloud-Based Stream Processing}\label{Sec:related:CBSP}
There are two general methodologies for processing streams in a cloud: using an operator-based DSMS, and discretizing stream inputs into mini-batches~\cite{zaharia2012discretized}. The former derives from traditional DSMSs described in Section~\ref{Sec:related:TradDSMS}, whereas the latter reduces stream processing to batch execution, explained in Section~\ref{Sec:related:rsCloud}. In general, mini-batch systems are optimized for throughput, at the expense of increased query response time, since each input must wait until a full batch is formed. While it is possible to minimize this extra latency by having extremely small batches, doing so would lead to high overhead, defeating the purpose. We focus on operator-based DSMSs since our target applications have real-time constraints, in which response time is key.

Two popular open source operator-based DSMSs are Storm~\cite{storm} and S4~\cite{neumeyer2010s4}. Their main difference is that Storm guarantees the correctness of its results (e.g., through its Trident component), while S4 does not. Both systems rely on manual configurations for resource scheduling. Hence, to avoid slow responses due to operator overloading, the user has to either overprovision resources to every operator, which is wasteful, or continuously tuning the system, which is infeasible for dynamic streams.

Many research prototypes of operator-based DSMSs are proposed, such as TimeStream~\cite{qian2013timestream}, which features efficient fault recovery, and Samza~\cite{samza}. None of these systems, however, addresses the resource scheduling problem. In the following we present DRS, the first effective resource scheduler for cloud-based operator DSMSs.

\section{Dynamic Resource Scheduling}\label{sec:drs}
Section~\ref{sec:drs:assumptions} clarifies assumptions in DRS. Section~\ref{sec:drs:model} presents the DRS performance model, which estimates query response time given a resource allocation scheme. Section~\ref{sec:drs:algo} describes the DRS dynamic resource scheduling algorithm. Table~\ref{Tab:notation} summarizes frequently used notations throughout the paper.
\begin{table}[htb]
\centering \caption{Table of notations.}
\label{Tab:notation}
\begin{tabular}{|c|p{2.5in}|}
\hline
    Symbol          &       Meaning\\
\hline
    $N$             & Total number of operators in an application\\
\hline
    $\lambda_i$     & Mean arrival rate of inputs to $i$-th operator\\
\hline
     $\lambda_0$    & Mean arrival rate of inputs to the application\\
\hline
    $\mu_i$         & Mean processing rate of inputs to $i$-th operator\\
\hline
    $k_i$           & \# of processors allocated to the $i$-th operator\\
\hline
    $\mathbf{k}$    & A Vector $(k_1,\ldots,k_N)$ containing all $k_i$s.\\
\hline
    $T_{\max}$      & Real-time constraint parameter: each input of the application is expected to be fully processed within $T_{\max}$ time.\\
\hline
    $K_{\max}$      & Resource constraint parameter: maximum number of available processors that can be allocated to the operators.\\
\hline
    $t$             & an input tuple to the streaming application \\
\hline
    $T$             & A random variable on the total sojourn time of a tuple $t$\\
\hline

\end{tabular}
\end{table}

\subsection{Assumptions}\label{sec:drs:assumptions}
We focus on stream analytics applications, which are usually memory-based and computation intensive. For such applications, \emph{processors} are the main type of resource, each of which contains a CPU (or one of its cores) and a certain amount of RAM. Disk space is not critical as streaming inputs are computed on-the-fly. Although networking delay can also affect query latency, we do not explicitly model it, because (a) it is often correlated with computational costs and (b) it can be affected by uncontrollable factors, such as other transmission-heavy applications on the same server or in the same subnetwork. Further, data centers today are increasingly equipped next-generation networking hardware that provide significantly higher bandwidth and lower latency, such as 10G Ethernet (e.g., in~\cite{soliman2014orca}) and InfiniBand (e.g., as argued in~\cite{mitchell2013using}), whose prices have been dropping rapidly. In contrast, processor speed in terms of CPU clock rate and RAM latency has stagnated in the past few years. Hence, we assume processors to be the bottleneck of the system, not network bandwidth.

For the ease of presentation, we further assume that all processors in the cloud have \emph{identical computational power}. Nevertheless, the proposed models and algorithms can also support settings with heterogeneous processors, and we explain how this is done whenever necessary. Meanwhile, we assume that \emph{load balancing} is achieved in every operator, i.e., each processor inside the same operator performs roughly equal amount of work. How to achieve load balancing is an orthogonal topic, and it is under active research, e.g.,~\cite{collins2009flexible,xing2005dynamic}. Under these assumptions, the processing speed of an operator depends mainly on the number of processors therein.

The goal of DRS is to fully process each input of the application in real time. Specifically, an input tuple to the application, e.g., a video frame in Figure~\ref{Fig:runExp1}, may lead to multiple intermediate results, e.g., features extracted by operator $A$, and objects recognized by operator $B$. We say an input tuple $t$ is fully processed, if and only if every intermediate result derived from $t$ has been processed by its corresponding operator. We use the term \emph{total sojourn time} to refer to the duration from the time that $t$ first arrives at the system to the time that $t$ is fully processed. Our goal is then to ensure that the expected total sojourn time of each input $t$ is no more than a user-specified duration, denoted by $T_{\max}$.

\subsection{Performance Model}\label{sec:drs:model}

Given an application's operator network, e.g., the one in Figure \ref{Fig:runExp2}, the current resource allocation and characteristics of the streaming data, the DRS performance model estimates the total sojourn time of an average input of the application, explained at the end of last subsection. The current resource allocation is represented by the number of processors assigned to each operator. Formally, we define $N$ as the number of operators in an application and a resource allocation is modeled by a vector $\mathbf{k} = (k_1, k_2, \dots, k_N)$, where $k_i (1\leq i\leq N)$ corresponds to the number of processors allocated to the $i$-th operator.

Regarding data characteristics, the important variables are the rate that tuples arrive at each operator, and how fast they can be processed by one processor. Networking delay is not explicitly expressed in our model, and we discuss this issue further at the end of this subsection. Note that our model assumes neither deterministic tuple arrival rates nor processing times; in other words, instantaneous arrival rates and processing times can fluctuate. On the other hand, in order to make the problem tractable, we do assume that the system remains in a relatively steady state during the span that DRS performs modeling and resource scheduling. This means that the average tuple arrival rate and processing time at each operator remains stable, and we obtain these quantities through the measurement module of the system, described in Section \ref{Sec:sysDesign}. Specifically, for the $i$-th operator ($1\leq i\leq N$), we use $\lambda_i$ to denote the mean arrival rate of its inputs, and $\mu_i$ to denote the mean processing rate of each of its processors. For instance, the case of $k_i=3$, $\lambda_i=10$ and $\mu_i=3$ means that on average, 10 tuples arrive at the $i$-th operator in unit time, and each of its 3 processors processes 3 tuples in unit time. For an operator with multiple input streams, i.e., join operators, $\lambda_i$ is the total arrival rates of all its input streams, and $\mu_i$ is the average processing rate of the operator, regardless of which input stream the tuple comes from.

Additionally, we define $\lambda_0$ as the mean arrival rate of inputs that flow into the application's operator network from outside of it. When there are clear ``source'' operators in the operator network whose inputs come entirely from outside the network, $\lambda_0$ is simply the total arrival rates of these sources. In general, however, there may not be a simple relationship between $\lambda_0$ and the set of $\lambda_i$'s, $1\leq i\leq N$. For example, in Figure~\ref{Fig:runExp2}, $\lambda_0$ is the arrival rate of tuples that come (from outside the system) to operator $A$; the input arrival rate $\lambda_A$ for $A$ on the other hand is the sum of $\lambda_0$ and the arrival rate of $A$'s other input stream, produced by operator $E$.

We use random variable $T$ to denote the total sojourn time of an input to the application. Our goal is to estimate $E[T]$, i.e., the expected value of $T$. The basic idea for estimating $E[T]$ is to model the system as an open queuing network (OQN)~\cite{bitran1996state}, and apply known results in queueing theory. In OQN, the total sojourn time of an input tuple $t$ is computed by summing up its total service time (i.e., total time spent on processing $t$ and intermediate results derived from $t$) and total queuing delay (total time that $t$ and its derived tuples wait in operator queues). This closely matches our setting. The challenge, however, is that there are numerous OQN models in the queuing literature, and selecting an appropriate one is non-trivial. On one hand, complex queuing network models generally do not have known solutions; among the ones that do, most have only numerical solutions (rather than analytical ones), which renders effective optimization hard; on the other hand, an overly simplified model may rely on strong assumptions, such as deterministic tuple arrival rates, which do not hold in our setting. After comparing various options and testing them through experiments, we chose to build our model based on a combination of one of Erlang's models~\cite{erlang1917solution,tijms1986stochastic} and the Jackson network~\cite{bitran1996state,jackson1963jobshop}. The former enables effective analysis of each individual operator, and the latter helps to aggregate these analyses to estimate $E[T]$ for the whole network. Our model has an analytical solution, and it involves only mild limitations, which will be discussed shortly.

%According to queuing theory, the sojourn time $T$ consists of two components: service time (i.e., total time that an input and intermediate results generated from it get processed at a processor) and queuing delay (total time that the input and its derivatives wait in operator input queues). As explained in Section~\ref{sec:drs:assumptions}, we do not explicitly consider networking delay in our model. Hence, the result always underestimates the true $E[T]$; nevertheless, as we show in Section~\ref{Sec:Exp}, our estimates strongly correlates with the true values.

We first focus on a single operator, say the $i$-th. We use $T_i$ to denote the time between the arrival of an input of the operator and the time when the operator finishes processing it. We model the operator as an M/M/$k_i$ system ~\cite{tijms1986stochastic}, where $k_i$ is the number of processors for operator $i$. According to the Erlang formula~\cite{tijms1986stochastic}, $E[T_i]$ is calculated by:
\begin{eqnarray}
\label{EQ:ETi}
E[T_i](k_i) =
    \left\{
        \begin{array}{ccl}
        \frac{\left(\frac{\lambda_i}{\mu_i}\right)^{k_i}\pi_0}{k_i!\left(1 - \frac{\lambda_i}{\mu_i k_i}\right)^2 \mu_i k_i} + \frac{1}{\mu_i}
        & \mbox{for} & k_i > \frac{\lambda_i}{\mu_i}; \\
        +\infty & \mbox{for} & k_i \leq \frac{\lambda_i}{\mu_i},
        \end{array}
    \right.
\end{eqnarray}
where $\pi_0$ is a normalization term, given by:

\begin{equation}
\pi_0 =
\left[\sum_{l=0}^{k_i-1} \frac{\left(\frac{\lambda_i}{\mu_i}\right)^l}{l!} + \frac{\left(\frac{\lambda_i}{\mu_i}\right)^{k_i}}{k_i!\left(1 - \frac{\lambda_i}{\mu_i k_i}\right)}\right]^{-1}.
\end{equation}

Intuitively, since new tuples arrive at an average rate $\lambda_i$, and each processor processes tuples at an average rate $\mu_i$, when $k_i \leq \frac{\lambda_i}{\mu_i}$, the processors cannot keep up with incoming tuples. Consequently, the number of tuples in the operator queue increases with time, leading to infinite queuing delay. When, $k_i > \frac{\lambda_i}{\mu_i} $, tuples are expected to be handled faster than they arrive. However, due to the randomness of the arrival and process rates, the queue may still grow when the arrival rate is temporarily higher than the processing rate. Clearly, the expected service time for each tuple is $\frac{1}{\mu_i}$. The expected queuing delay is captured by the complicated term in Equation~(\ref{EQ:ETi}).

Next we aggregate all $E[T_i]$'s to obtain an estimate of $E[T]$ for the entire operator network. According to the theory of Jackson networks~\cite{bitran1996state,jackson1963jobshop}, $E[T]$ is computed by a weighted average of the $E[T_i]$'s:
\begin{equation}
E[T](\mathbf{k}) = E[T](k_1, k_2, \dots, k_N) = \frac{1}{\lambda_0}\sum_{i = 1}^N \lambda_iE[T_i](k_i)\label{EQ:ET}.
\end{equation}

This completes the DRS performance model. Since our model relies on Erlang's formula and Jackson open queueing network, it inherits two limitations. First, the model implicitly assumes that both the inter-arrival times of external tuples (that come from outside the system) and the service time of the operator are {\it i.i.d.} samples from random variables following the exponential distribution. Second, Jackson network does not explicitly model pipelining between different operators. Hence, our model may give an inaccurate estimate of $E[T]$, when the service time or tuple arrival distribution deviates significantly from the expected exponential distribution, or when pipelining affects total processing time considerably. In the meantime, our model does not explicitly consider networking costs, due to the fact that measuring the networking delay between two nodes requires complex inter-node protocols, e.g., for clock synchronization, which can be prohibitively expensive in a real-time application. Therefore, when networking delay becomes a dominant factor in the total sojourn time of an average input, our model tends to produce an underestimation of the true result. Nevertheless, as we show in the experiments, the value of $E[T]$ predicted by our model is sufficiently accurate, when the underlying application is computation intensive, which is one important assumption made in Section \ref{sec:drs:assumptions}. Further, even when the prediction is inaccurate, it is still strongly correlated with the exact value of $E[T]$, meaning that DRS remains capable of identifying the best resource allocation with the predicted value. In the rest of the section, we show how DRS schedules resources based on the performance model.

\subsection{Scheduling Algorithm}\label{sec:drs:algo}
In a nutshell, DRS (a) monitors the current performance of the system (more details in Section~\ref{Sec:sysDesign}), (b) checks whether the performance falls (or is about to fall) under the real-time constraint, or when the system can fulfil the constraint with less resources, and (c) reschedules resources when (b) returns a positive result. The main challenge lies in (b), which needs to answer two questions, including \emph{how many} processors are needed to fulfil the real-time requirement, and \emph{where} to place them in the operator network. We first focus on the latter question. Specifically, given a number (say, $K_{\max}$) of processors, we are to find an optimal assignment of these processors to the operators of the application that obtains the minimum expected total sojourn time. The problem can be mathematically formalized as follows:
\begin{equation}
\label{opt1}
\begin{split}
\min\limits_{\mathbf{k}}\;\;& E[T](\mathbf{k})\\
\textrm{s.t.}\;\;& \sum_{i=1}^N k_i \leq K_{\max},\;k_i\;\textrm{is interger}, i = 1, 2, \dots, N;
\end{split}
\end{equation}

A naive solution to the above optimization problem is to view it as an integer program, and apply a standard solver. However, current integer programming solvers are prohibitively slow, especially considering that DRS itself has to run in real time. In the following we describe a novel algorithm that solves Program~(\ref{opt1}) with negligible cost.

The key property used in the proposed algorithm is that $E[T_i](k_i)$, defined in Equation~(\ref{EQ:ETi}), is a convex function of $k_i$, the number of processors assigned to the $i$-th operator. This property has already been proved in~\cite{bitran1996state}. It follows from the convexity of $E[T_i](k_i)$ implies that marginal benefit for incrementing $k_i$ drops monotonously as $k_i$ becomes larger. Formally, for all $k'_i > k_i$, we have:
\begin{equation}
\label{EQ:ETiConvex}
E[T_i](k_i) - E[T_i](k_i + 1) > E[T_i](k'_i) - E[T_i](k'_i + 1)
\end{equation}

Now observe from Equation~(\ref{EQ:ET}) that $E[T]$ is a weighted sum of the $E[T_i]$'s, and each weight $\lambda_i$ is independent of the value of $k_i$. Hence, $E[T]$ is also a convex function of the $k_i$'s, meaning that incrementing each $k_i$ also has diminishing marginal benefit with respect to $E[T]$. Based on this observation, we design a greedy algorithm, listed in Algorithm~\ref{Alg:BoxmaT}. The idea is to start from the smallest possible value of each $k_i$ (lines 1-4) and iteratively add one processor to the operator that leads to the largest decrease in $E[T]$ (lines 8-15). According to Equation~(\ref{EQ:ETi}), each $k_i$ must be larger than $\frac{\lambda_i}{\mu_i}$, since otherwise, $E[T_i](k_i)$ becomes infinitely large, leading to infinity on $E[T]$ as well.
\begin{algorithm}
\caption{$AssignProcessors$}
\label{Alg:BoxmaT}
\begin{algorithmic}[1]
\REQUIRE $K_{\max}$, $\lambda_0$, $\{\lambda_i, i = 1, \dots, N\}$, $\{\mu_i, i = 1, \dots, N\}$.
\ENSURE $\mathbf{k} = (k_1, k_2, \dots, k_N)$
\FORALL {$i \gets 1, \dots, N$}
    \STATE $k_i \gets \left\lceil\frac{\lambda_i}{\mu_i}\right\rceil$ \COMMENT{Initialize each $k_i$}
\ENDFOR
\IF {$\sum_{i=1}^N k_i > K_{\max}$}
\STATE throw an exception saying that the number of processors are not sufficient for the application.
\ENDIF
\WHILE {$\sum_{i=1}^N k_i < K_{\max}$}
    %\STATE \COMMENT{In each round, find the operator $j$ with the maximum marginal benefit.}
    \FORALL {$i \gets 1, \dots, N$}
        \STATE $\delta_i\gets \lambda_i\cdot\Big[E[T_i](k_i) - E[T_i](k_i + 1)\Big]$
    \ENDFOR
    \STATE \COMMENT{find the operator with the largest marginal benefit.}
    \STATE $j\gets\arg\max_i \delta_i$
    \STATE $k_j\gets k_j + 1$
\ENDWHILE
\RETURN $\mathbf{k} = (k_1, k_2, \dots, k_N)$
\end{algorithmic}
\end{algorithm}

Since $E[T]$ is convex, the above greedy algorithm always finds the optimal solution, similar to the case of the server reallocation problem~\cite{boxma1990machine}. This is restated as follows:
\begin{theorem}\label{theroem1}
Algorithm \ref{Alg:BoxmaT} always returns exact optimal solution to Program~\ref{opt1}.
\end{theorem}
The proof is given in Appendix \ref{appendix:theroemProof}.

Next we focus on the question on how to determine the minimum number of processors that are expected to achieve real-time processing, i.e., the expected total sojourn time $E[T]$ is no larger than a user-defined threshold $T_{\max}$. This can be modeled with the following optimization problem.
\begin{eqnarray}
\label{opt2}
\begin{split}
\min\limits_{\mathbf{k}}\;\;& \sum_{i=1}^{N} k_i,\\
\textrm{s.t.}\;\;& E[T](\mathbf{k}) \leq T_{\max},\;k_i\;\textrm{is interger}, i = 1, 2, \dots, N;
\end{split}
\end{eqnarray}

Similar to Program~(\ref{opt1}), both constraints and objective of Program~(\ref{opt2}) are convex in terms of $\mathbf{k}$. Hence, we solve Program~(\ref{opt2}) with a greedy strategy similar to Algorithm~\ref{Alg:BoxmaT}. Specifically, we start by initializing each $k_i$ with minimal requirement, as in lines 1-4 of Algorithm \ref{Alg:BoxmaT}. The algorithm repeatedly adds one processor to the operator with the maximum marginal benefit as in lines 8-15 of Algorithm \ref{Alg:BoxmaT}, until $E[T]$ is no larger than $T_{\max}$. We omit the proof of correctness for this algorithm, since it is nearly identical to that of Algorithm~\ref{Alg:BoxmaT}.

In practice, the solution of Program~(\ref{opt2}) may not give us the precise amount of resources necessary for meeting the real-time requirement at all times, for two reasons. First, the total sojourn time can be different for every input, and $E[T]$ is merely its expected value. Second, the performance model described in Section~\ref{sec:drs:model} outputs only an estimate of $E[T]$, rather than its precise value. To address this problem, DRS starts with the number of processors suggested by the solution of Program~(\ref{opt2}), monitors the actual total sojourn time $E[\hat{T}]$, and continuously adjusts the number of processors based on the measured value of $E[\hat{T}]$. In next section, we discuss the system design and implementation issues with DRS.

\section{System Design}\label{Sec:sysDesign}

\begin{figure}[htb]
    \centering
    \vspace{-5pt}
    \includegraphics[width=3.2in]{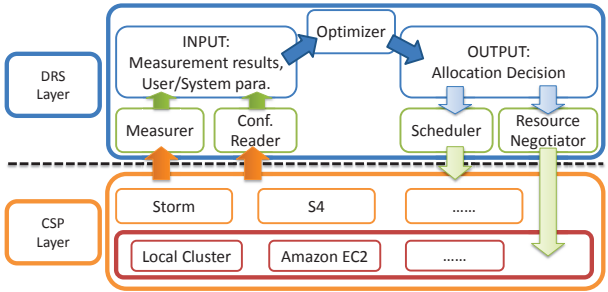}
    \caption{The architecture overview}
    \label{Fig:ERSOverview}
    \vspace{-5pt}
\end{figure}

An overview of the system architecture is presented in Figure~\ref{Fig:ERSOverview}, which generally consists of two layers, the DRS layer and the CSP (cloud-based streaming processing) layer. Specifically, DRS layer is responsible for performance measurement, resource scheduling and resource allocation control, while CSP layer contains the primitive streaming processing logic, e.g. running instances of \emph{Storm}~\cite{storm} and \emph{S4}~\cite{neumeyer2010s4}, and cloud-based resource pool service, e.g. YARN~\cite{yarn} and Amazon EC2.

While the core of DRS layer is responsible to the optimization of resource scheduling based on the model derived in the previous section, the system to support such functionality is not that straightforward to build. Given the heterogeneous underlying infrastructure and the complicated streaming processing applications running on the CSP layer, it is crucial to collect the accurate metrics from the infrastructure, aggregate the statistics, make online decisions and control the resource allocation in an efficient manner.

To seamlessly combine the optimization model and the concrete streaming processing system, we build a number of independent functional modules, which bridge the gap between the physical infrastructure and abstract performance model. As is shown in Figure~\ref{Fig:ERSOverview}, on the input side of the optimizer component, we have \emph{measurer} module and \emph{configuration reader} module, which generate the statistics needed by the optimizer based on the data/control flow from CSP layer. On the output end of the workflow diagram, the \emph{scheduler} module and \emph{resource negotiator} module transform the decisions of the optimizer into executable commands for different streaming processing platforms and resource pools.
The technical details and key features of the modules are discussed in Appendix \ref{appendix:drsModules}.

\section{Empirical Studies}\label{Sec:Exp}

%Section~\ref{sec:exp:platform} describes how we implement DRS inside Storm, a popular cloud-based operator DSMS. Section~\ref{sec:exp:app} present the real applications and data used to test DRS. Section~\ref{sec:exp:setup} details experimental setup. Section~\ref{sec:exp:results} provides the evaluation results.

To test the effectiveness of DRS, we have implemented it\footnote{The source code is available online: \url{https://github.com/ADSC-Cloud/resa/}} and integrated it into \emph{Storm} \cite{storm}, which provides the underlying CSP layer.
%In this subsection, we briefly overview important concepts and architectural aspects of Storm, and describe the implementation of the measurer, scheduler and resource negotiator modules of DRS in Storm.
The overview of the important concepts and architectural aspects of Storm, and the description of how we implement the measurer, scheduler and resource negotiator modules of DRS in Storm are provided in Appendix \ref{appendix:platform}.

\subsection{Testing Applications}\label{sec:exp:app}

We implement two real-time stream analytics applications: video logo detection (VLD) and frequent pattern detection (FPD) from different domains.

%\subsubsection{Logo Detection from a Video Stream}\label{sec:exp:app:logo}

\noindent\textbf{Logo Detection from a Video Stream.} Given a set of query logo images, the logo detection application identifies these images from the input video stream. Although much work has been done to improve the accuracy and efficiency of VLD, performing it in real time remains a major challenge, due to the high computational complexity.

\begin{figure}[htb]
    \centering
    \vspace{-5pt}
    \includegraphics[width=2.6in]{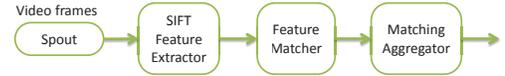}
    \caption{The topology of real-time video logo detection application.}
    \vspace{-5pt}
    \label{Fig:logodet}
\end{figure}

Figure~\ref{Fig:logodet} illustrates the topology of the real-time VLD application, which is a chain of operators containing a spout, a feature extractor, a feature matcher, and an aggregator. The spout extracts frames from the raw video stream. The output rate of frames may vary from time to time due to the generation algorithm and the original video contents. We employ scale-invariant feature transform (SIFT)~\cite{lindeberg2012scale} algorithm to extract features from each frame. This step is time-consuming, involving convolutions on the 2-dimensional image space. Moreover, the number of result SIFT features may vary dramatically on different frames, causing significant variance on the computation overhead over time. The feature matcher measures $L_2$ distance between its input SIFT features to those pre-generated logo features, and outputs matching pairs with distance lower than a pre-defined threshold. Finally, the aggreagator judges whether a logo appears in a video frame by aggregating all input matching feature pairs, i.e., if the number of matched features in a video frame exceeds a threshold, the logo is considered to appear in the frame.
%Note that it is possible to find multiple matches with a single query log image on one frame.

%\subsubsection{Frequent Pattern Detection over a Tweet Stream}\label{sec:exp:app:freq}

\noindent\textbf{Frequent Pattern Detection over a Microblog Stream.} This application maintains the frequent patterns~\cite{AS94} on a sliding window over a microblog stream from Twitter. For each input sentence, we append an additional label ``+/-'', indicating it is entering/leaving the dedicated window. Given a set of input item groups in the sliding window and a threshold, we define a maximal frequent pattern (MFP) to be the itemset satisfying: (a) the number of item groups containing this itemset, called its occurrence count, is above the threshold; and (b) the occurrence count of any of its superset is below the threshold.

\begin{figure}[htb]
    \centering
    \vspace{-5pt}
    \includegraphics[width=2.6in]{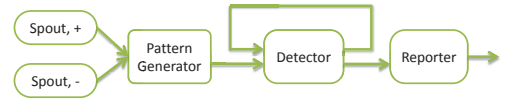}
    \caption{The topology of the stream frequent pattern detection application.}
    \vspace{-5pt}
    \label{Fig:fpdet}
\end{figure}

Figure~\ref{Fig:fpdet} illustrates the operator topology. There are two spouts, which generate an event tuple as an itemset enters/leaves the current processing window, respectively. The pattern generator generates candidate patterns, i.e., itemsets. These candidates include an exponential number of possible non-empty combinations of items. Hence, its computation varies, according to the number of items in recent transactions.

The detector maintains the state records containing (a) the occurrence counts and (b) MFP indicator, of all the candidate itemsets. When a state change happens to some itemset, e.g., from MFP to non-MFP, the detector outputs a notification to the reporter, and also to itself through the loop back link.
%The reason for this loop is that each processor in the detector maintains a portion of the state records; because a state change can affect the states of other itemsets stored at a different processor, the loop ensures that the state change notifications be sent to all the instances.
Since (a) each processor in the detector maintains only a portion of the state records; and (b) a state change can affect the states of other itemsets stored at a different processor, the loop ensures that the state change notifications be sent to all the instances.
Finally, the reporter presents the updates of the detection results to the user. In our implementation, the reporter simply write its inputs to an HDFS file. % David: the need for a loop seems a bit artificial.

\subsection{Experiment Setup}\label{sec:exp:setup}

The experiments were run on a cluster of 6 Ubuntu Linux machines interconnected by a LAN switch. Each machine is equipped with an Intel quad-core CPU \@3.4GHz and 8GB of RAM.
Following common configurations of Storm, we allocated one machine to host the Nimbus and the Zookeeper Server; the remaining 5 machines host executors for the experimental applications. We also configured each of these 5 machines so that one machine can host at most 5 executors. The main purpose of this constraint is to mitigate the interference caused by other executors running on the same machine, and the resource contention due to the over-allocation of executors on a single machine. As a result, there are 25 executors in total.

For both applications, namely video logo detection (VLD) and frequent pattern detection (FPD), we allocated two executors as spouts, and one executor for DRS. The remaining $25-3=22$ executors are used as bolts, i.e., $K_{\max} = 22$.
For VLD, the input data are a series of videos clips of the soccer games, and we selected 16 logos as the detection targets. The frame rate simulates a typical Internet video experience, which is uniformly distributed in the interval $[1, 25]$ with a mean of 13 frames/second.
For FPD, we use a real dataset containing 28,688,584 tweets from 2,168,939 users collected from Oct. 2006 to Nov. 2009. We set the sliding window to 50,000 tweets, and simulated the arrival of tweets to the topology following the Poisson process with an average arrival rate of 320 tweets per second.

\subsection{Experimental Results}\label{sec:exp:results}

For both applications, we run two sets of experiments: (a) with re-balancing\footnote{This is a term used by Storm, and it has the same meaning as re-scheduling.} disabled, i.e., we keep DRS running passively, meaning that it continues to monitor the system performance and recommend new (if better) resource allocation configurations, but does not perform re-scheduling; (b) re-balancing is disabled at the beginning, and then enabled at a later time. These experiments aim to test the quality of the performance model and evaluate the effectiveness of the resource scheduling algorithm of DRS.

\begin{figure}[htb]
    \centering
    \vspace{-5pt}
    \includegraphics[width=2.6in]{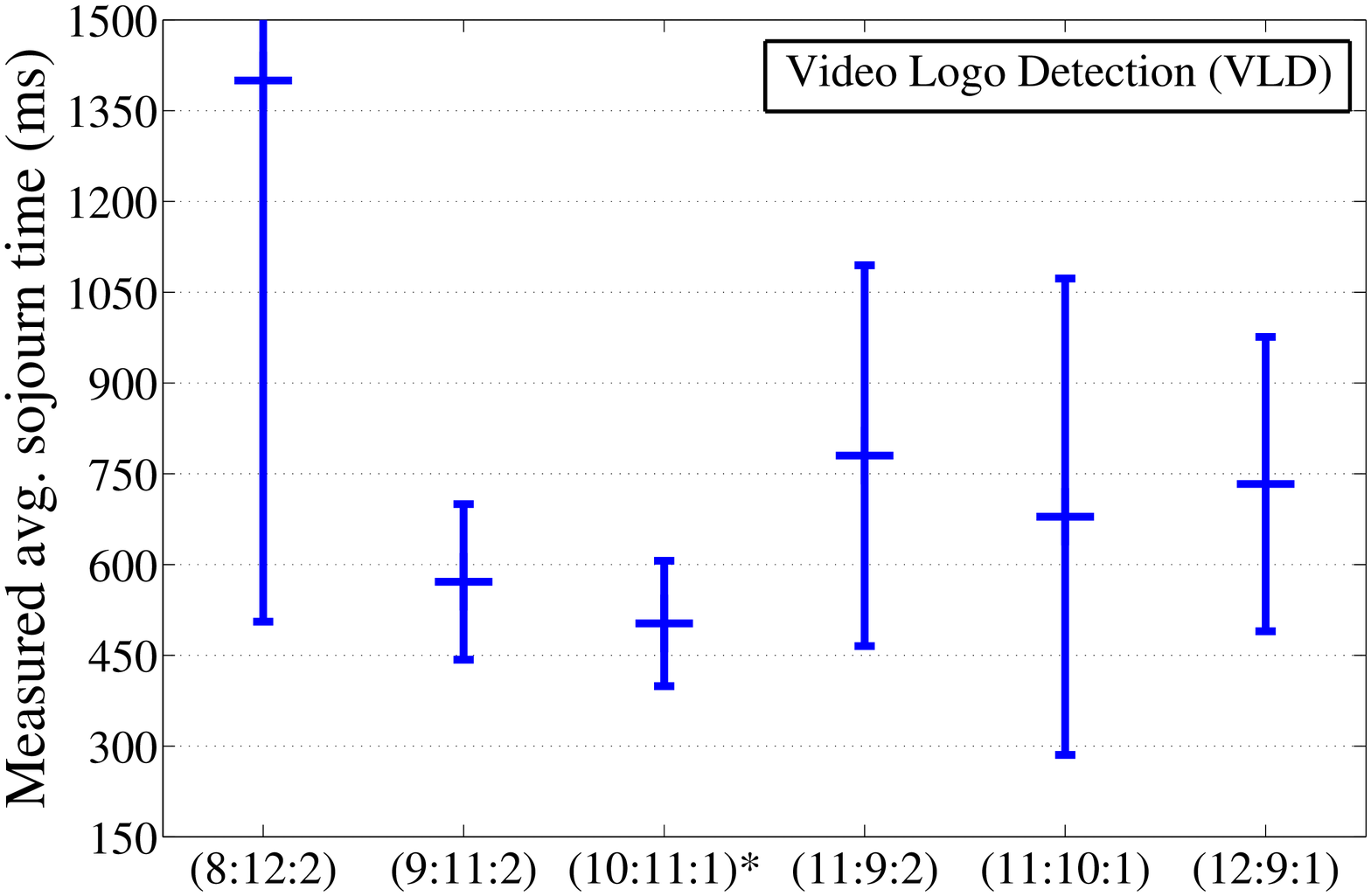}
    \includegraphics[width=2.6in]{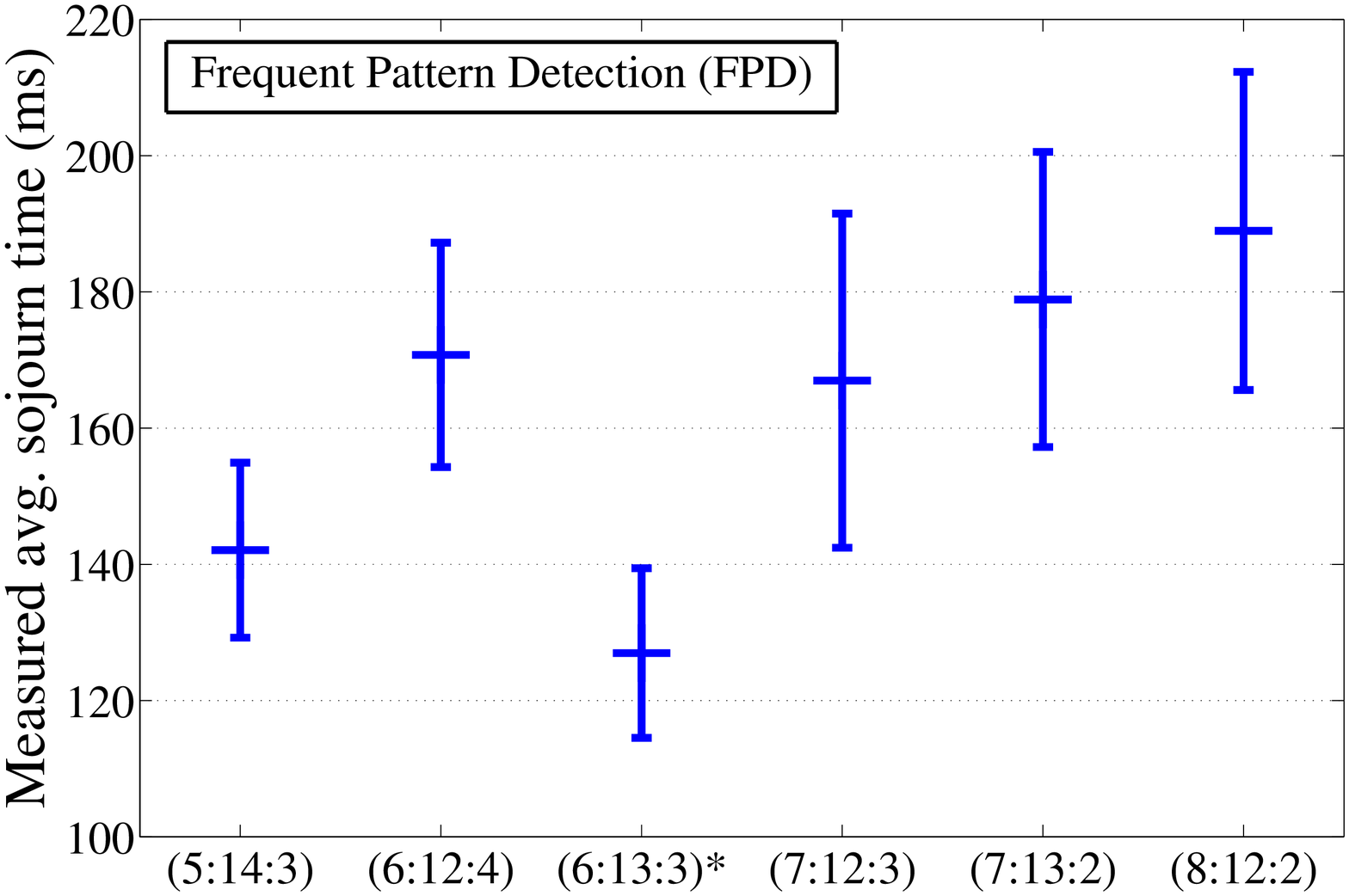}
    \vspace{-5pt}
    \caption{The mean and standard deviation of the complete sojourn times under different resource configurations with re-balancing disabled, where the configurations with ``$\ast$'' are the recommended allocations by the passively running DRS.}
    \label{Fig:logdetMfpBox}
\end{figure}

\noindent\textbf{Experiments with re-balancing disabled. } In this set, each experiment lasts for 10 minutes. Figure~\ref{Fig:logdetMfpBox} shows the mean and standard deviation of the total sojourn times under 6 different allocations for each application. The x-axis ($x_1$:$x_2$:$x_3$) denotes an resource configuration (in a partial order of $x_1, x_2, x_3$), where $x_1, x_2, x_3$ are the number of executors allocated to the operators \emph{SIFT Feature Extractor}, \emph{Feature matcher}, and \emph{Matching aggregator} in Figure~\ref{Fig:logodet}, or the \emph{Pattern generator}, \emph{detector}, and \emph{reporter} in Figure~\ref{Fig:fpdet}. The two configurations with ``$\ast$'', (10:11:1) for VLD and (6:13:3) for FPD are the recommended allocations by the passively running DRS.

From Figure~\ref{Fig:logdetMfpBox}, we make the following observations. The resource configurations (10:11:1) for VLD and (6:13:3) for FPD, achieve the best performance according to the measured average sojourn time. This turns out to be consistent with the recommendations provided by the passively running DRS, which validates the accuracy and effectiveness of our DRS performance model and resource scheduling algorithm.

In particular, these two configurations not only obtain the smallest average sojourn times, but also the minimum standard deviation, meaning that these two allocations lead to the smallest performance oscillations. Different configurations, including the 5 closest ones in terms of the $L_1$ distance (i.e., the remaining 5 in the experiment) to the best configurations (10:11:1) for VLD and (6:13:3) for FPD, all exhibit considerably worse performance. These results demonstrate that it is not trivial to find the optimal resource allocation especially when the application topology becomes more complicated (e.g. more than three bolt operators), and hence reveal the importance and usefulness of the DRS.

%The suggested allocations by the passively running DRS for both DLD and FPD are consistent to the configurations achieving the best
%Since DRS still runs passively in these experiments, we examined its suggested resource allocation configurations. It turns out that DRS recommends exactly the same configuration (10:11:1) as Figure ~\ref{Fig:logdetMfpBox} suggests. This validates the effectiveness of our the DRS performance model and resource scheduling algorithm.

\begin{figure}[htb]
    \centering
    \vspace{-5pt}
    \includegraphics[width=2.6in]{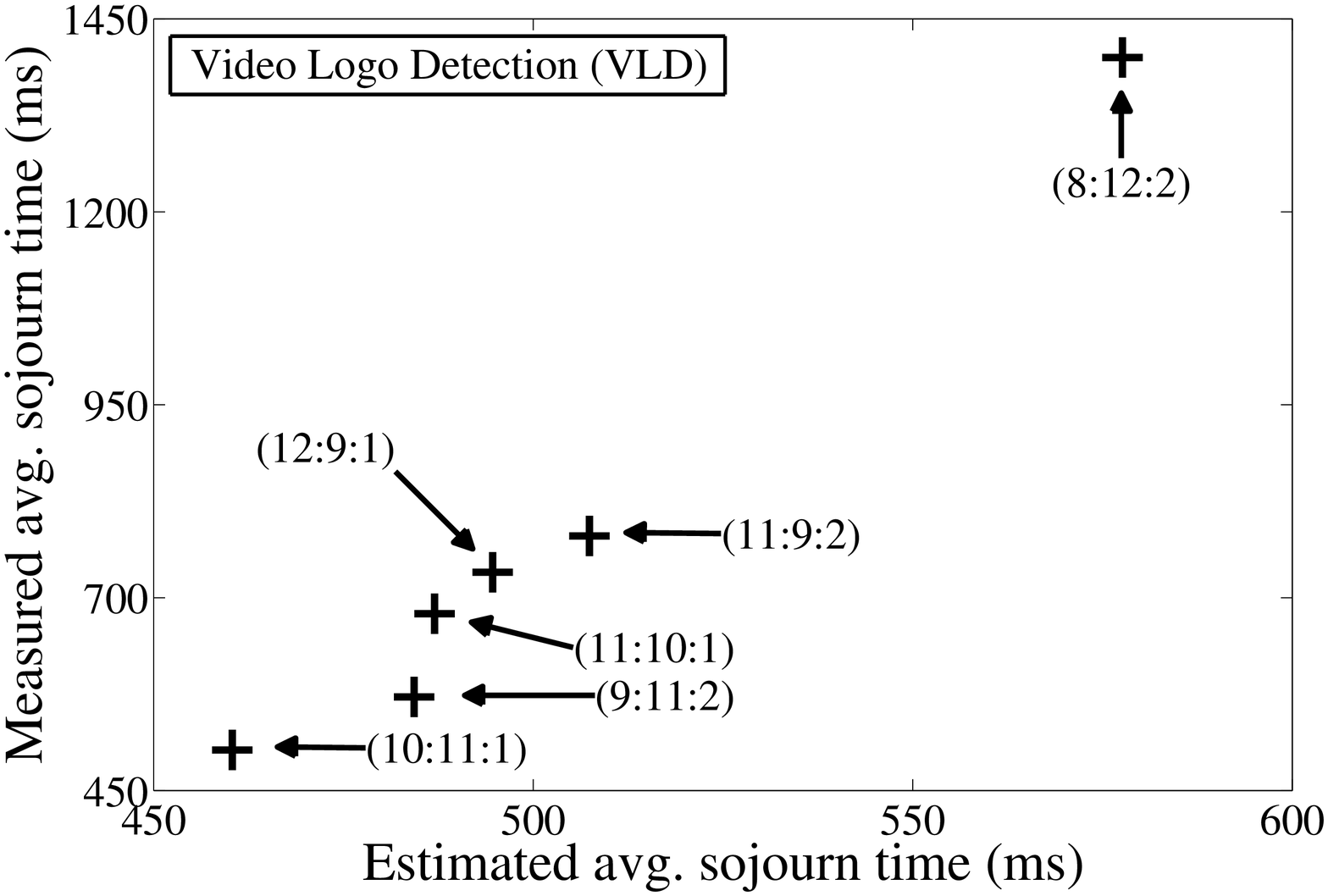}
    \includegraphics[width=2.6in]{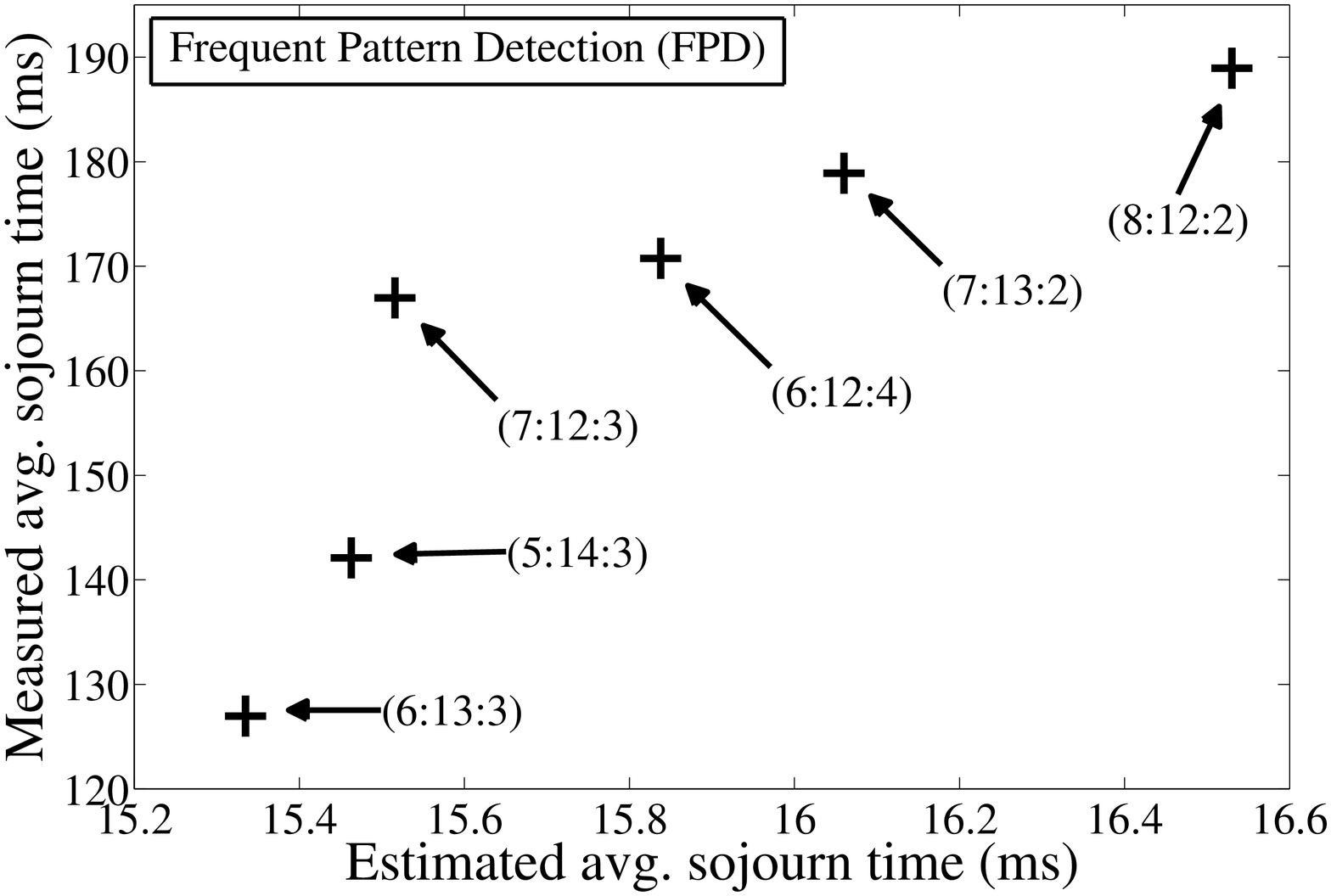}
    \vspace{-5pt}
    \caption{Comparing average sojourn time estimated by the model and measured in the experiment.}
    \label{Fig:logdetMfpScatter}
\end{figure}

To take a close look at how DRS provides resource configuration recommendations, correctly, Figure~\ref{Fig:logdetMfpScatter} shows the relationship between the measured average sojourn times and the estimated average sojourn time, which is derived by the performance model described in Section~\ref{sec:drs:model}, of the six resource allocation configurations for both VLD and FPD, with re-balancing disabled.

%Figure~\ref{Fig:logdetMfpScatter} shows the relationship between the average sojourn time measured in the experiment, and that estimated by the performance model, described in Section 3.2, which is run passively within the disabled DRS.

%If we connect the points in Figure~\ref{Fig:logdetMfpScatter}, the resulting curves for both applications are strictly monotonic,
As shown in Figure~\ref{Fig:logdetMfpScatter}, the points representing the measured and the estimated average sojourn time are showing the strictly monotonicity,
which signifies that the performance model is capable of suggesting the best resource allocation configuration.  Moreover, the performance model outputs accurate estimates for VLD;
%in particular, the estimates are only slightly below their corresponding measured values, %
though, with some slight underestimation comparing to the measured values,
which is expected, as our model does not consider network overhead. It is worth noting that the estimates are accurate even though the underlying conditions are not satisfied for the Jackson network theory and Erlang model.
%In particular,
For example, the frame rate is uniformly (rather than exponential as required) distributed. Meanwhile, the operator input queues do not follow strict FIFO rule; instead, tuples are hashed to processors. Different operators are also run in parallel, which leads to pipelining. The model is clearly robust to these variations of the conditions.

%For FPD, the performance model outputs estimates that are much lower than the actual sojourn times.
%
For FPD, the estimated sojourn times show larger deviations to the measured ones.
This is mainly because the model does not consider network transmission cost, which takes a dominant portion of the total query latency in this particular application.
%
%In other words, the FPD application is not the type of computation intensive application that we focus on.
In other words, the FPD is de facto the type of data intensive rather than the computation intensive application that we focus on.
Nevertheless, our model still correctly indicates the relative order of the performance of different resource allocation configurations. Meanwhile, since the estimates are strongly correlated with the true values, a polynomial regression
%over the estimated and true latency values can be used
can be used straightforwardly
to make accurate predictions of the true latency value given the estimated one.

\begin{figure}[htb]
    \centering
    \vspace{-5pt}
    \includegraphics[width=2.4in]{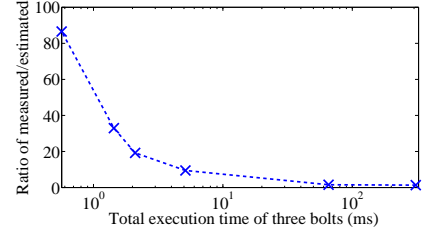}
    \vspace{-5pt}
    \caption{The degree of underestimation (the ratio of the measured to the estimated average sojourn time) v.s. the total CPU time of the three bolts of the synthetic chain topology}
    \label{Fig:3bolt}
\end{figure}

To further validate the above explanation, we carried out a separate experiment over a synthetic topology with a simple chain of three operators. Each operator simply performs some computations (such as empty for-loops) with varying load (e.g., number of loops). We used 30 executors running on 6 physical machines, connected in the same subnetwork.
%Then, we vary the CPU load of the operators. The results are reported in Figure~\ref{Fig:3bolt}, which shows the difference between estimation and actual latency.
The results are reported in Figure~\ref{Fig:3bolt}.

As shown in Figure~\ref{Fig:3bolt},
We tried 6 different workloads in terms of total CPU time (excluding the queue time) of the three bolts, from 0.567 millisecond, to 309.1 milliseconds (x-axis, log-scale), and the y-axis shows the ratio of the measured average sojourn time to the estimated value.
%It shows that the larger the computing workloads, more accurate the model estimates (ratio closer to 1).
It shows a clear decreasing trend of the degree of underestimation (ratio of the measured to the estimated average sojourn time) as the total CPU time of the three bolts increases.

\noindent\textbf{Experiments with re-balancing disabled first and then enabled.}
%Next, we run a group of experiments with re-allocation functionality enabled in order to see the dynamics of the allocation and their performance. In order to have a clear view of the performance (in terms of the average sojourn time) across re-scheduling events, we do not allow scheduler to apply the re-scheduling plan in the first 9 minutes in the experiments.
In this set of experiments, we investigate the performance of the real re-scheduling operation activated and executed by the DRS when it detects the non-optimal resource allocation configurations. For each experiment, it lasts for 27 minutes and the re-balancing function is disabled from the beginning till the end of the 13th minute, and becomes enabled afterwards. In this way, we are able to have a clear view of the performance (in terms of the average sojourn time) across the re-scheduling events.

\begin{figure}[htb]
    \centering
\vspace{-5pt}
    \includegraphics[width=2.6in]{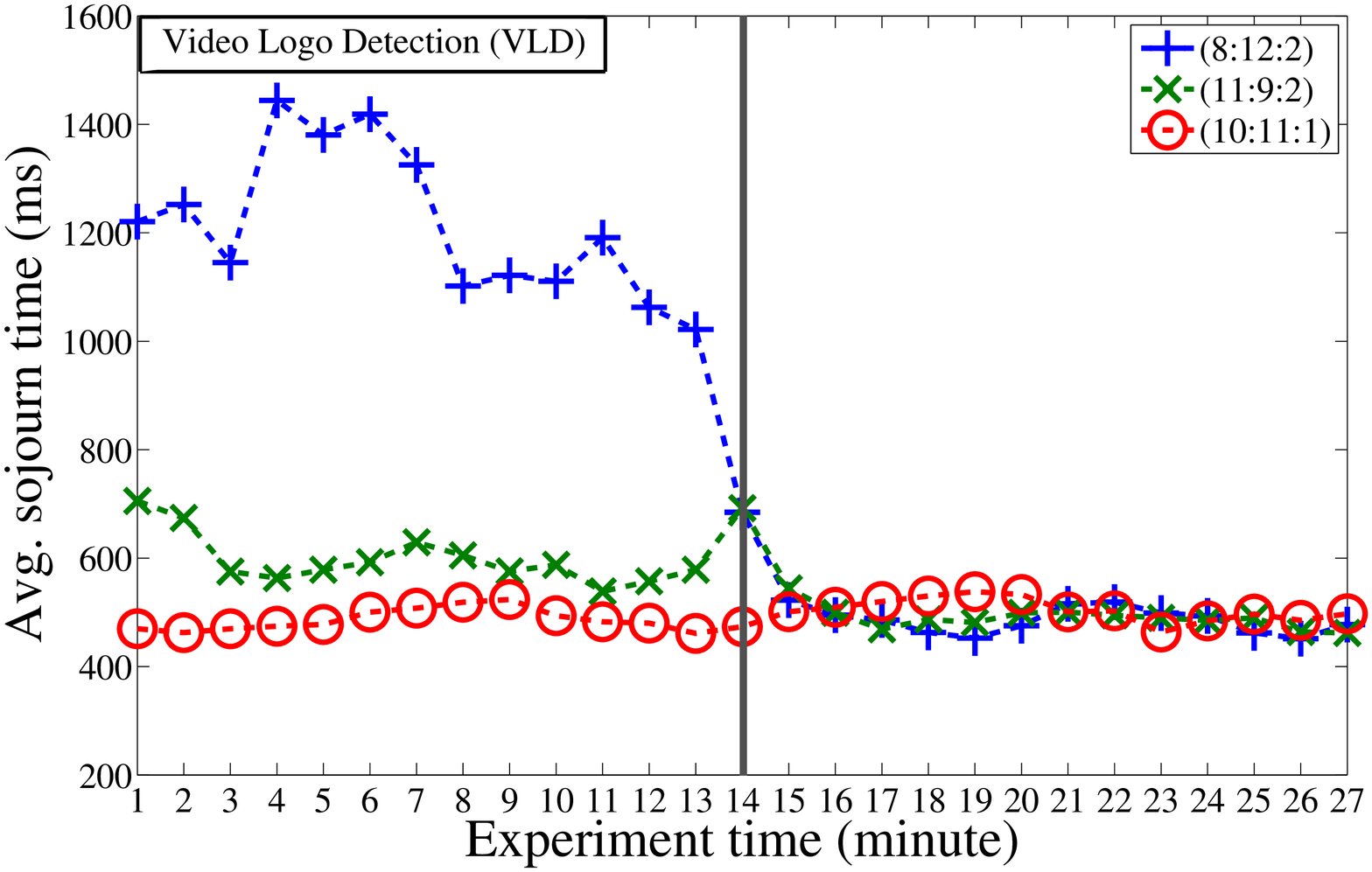}
    \includegraphics[width=2.6in]{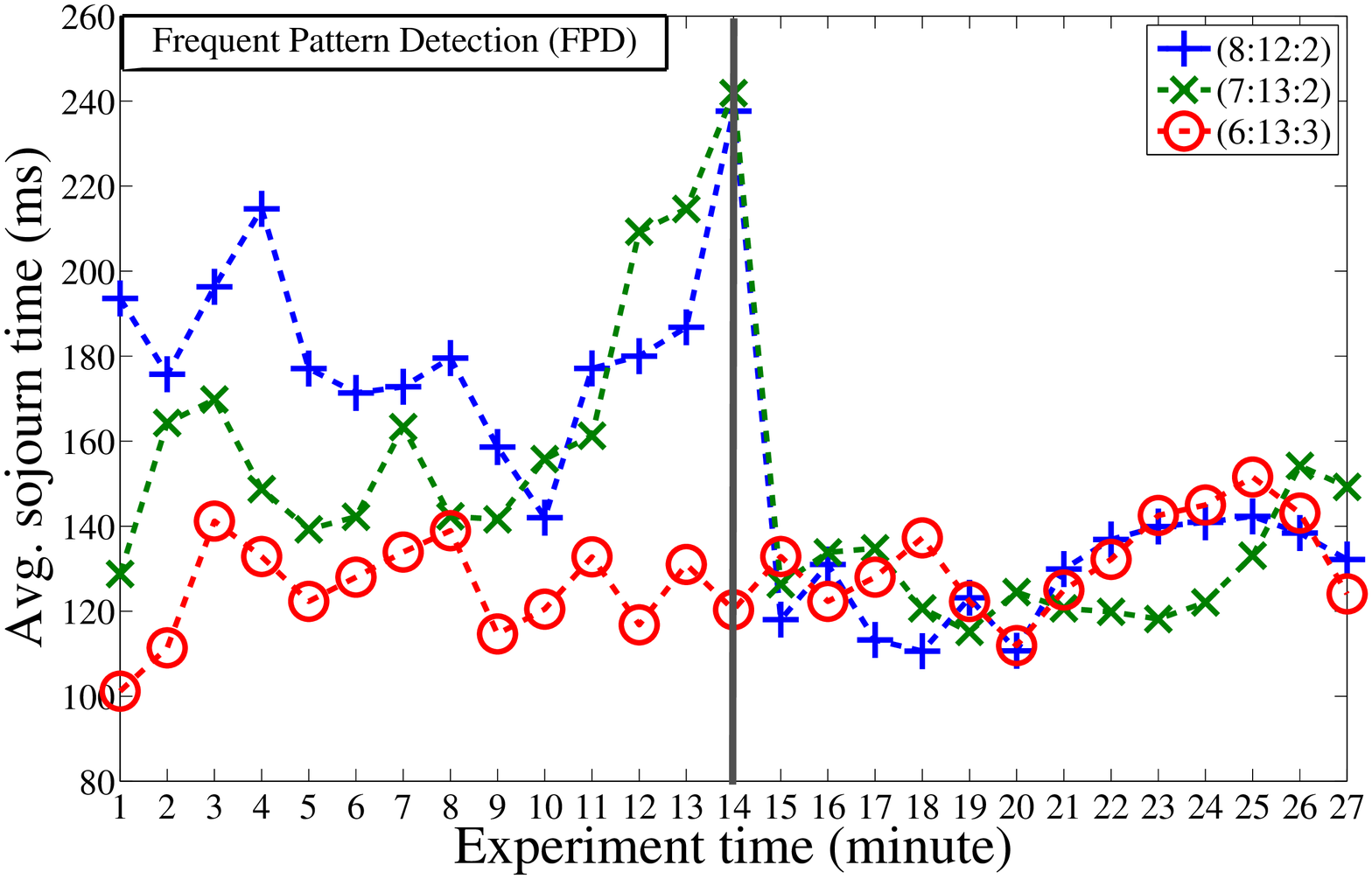}
\vspace{-5pt}
    \caption{The average sojourn times of three different allocations in the initial state for each application, where re-balancing function is disabled from the beginning till the end of the 13th minute, and becoming and keeping enabled since the 14th minute.}
    \label{Fig:logdetMfpReb}
\end{figure}
Figure~\ref{Fig:logdetMfpReb} shows three curves for each of the applications. In particular, each curve represents an initial allocation. For both applications, the two curves initially with the non-optimal allocations, experience the re-scheduling events at the 14th minute, while the one with the optimal allocation as its initial state does not. From Figure~\ref{Fig:logdetMfpReb}, we can see that optimizer triggers the re-scheduling action as early as possible, which responds quickly to the less promising resource scheduling plan. After the re-scheduling, all the curves with different initial allocations, were scheduled with the unique optimal solution. This statement is supported by two facts: a) from Figure~\ref{Fig:logdetMfpReb}, after the 14th minute, all the three curves have the similar average sojourn time, and similar performance trends. Especially for the two curves that experience the re-allocation event, it shows a clear decrease in the average sojourn time; b) the plans kept in the log files further verify this observation.

Another observation we make, according to the four curves experiencing the re-scheduling events shown in Figure~\ref{Fig:logdetMfpReb} - (8:12:2) and (11:9:2) of VLD and (8:12:2) and (7:13:2) of FPD - is that our improved version of re-balancing mechanism led to remarkably low cost, i.e. a neglectable increment in the average sojourn time within the 14th minute only. Besides, the whole re-balancing process of ours only takes a few seconds, comparing to the 1-2 minutes taken by Storm's default version.
%It is worth noting that executing the default version of the ``re-balancing'' function provided by the latest Storm release will cause serious performance degradation (i.e., the average sojourn time increases dramatically and lasts for a long duration, say 1 - 2 minutes. This is mainly because the default version of ``re-balancing'' function simply shuts down all the worker processes (Java virtual machines, JVMs), and then re-starts them under the new allocation configurations.

\begin{figure}[htb]
    \centering
\vspace{-5pt}
    \includegraphics[width=2.6in]{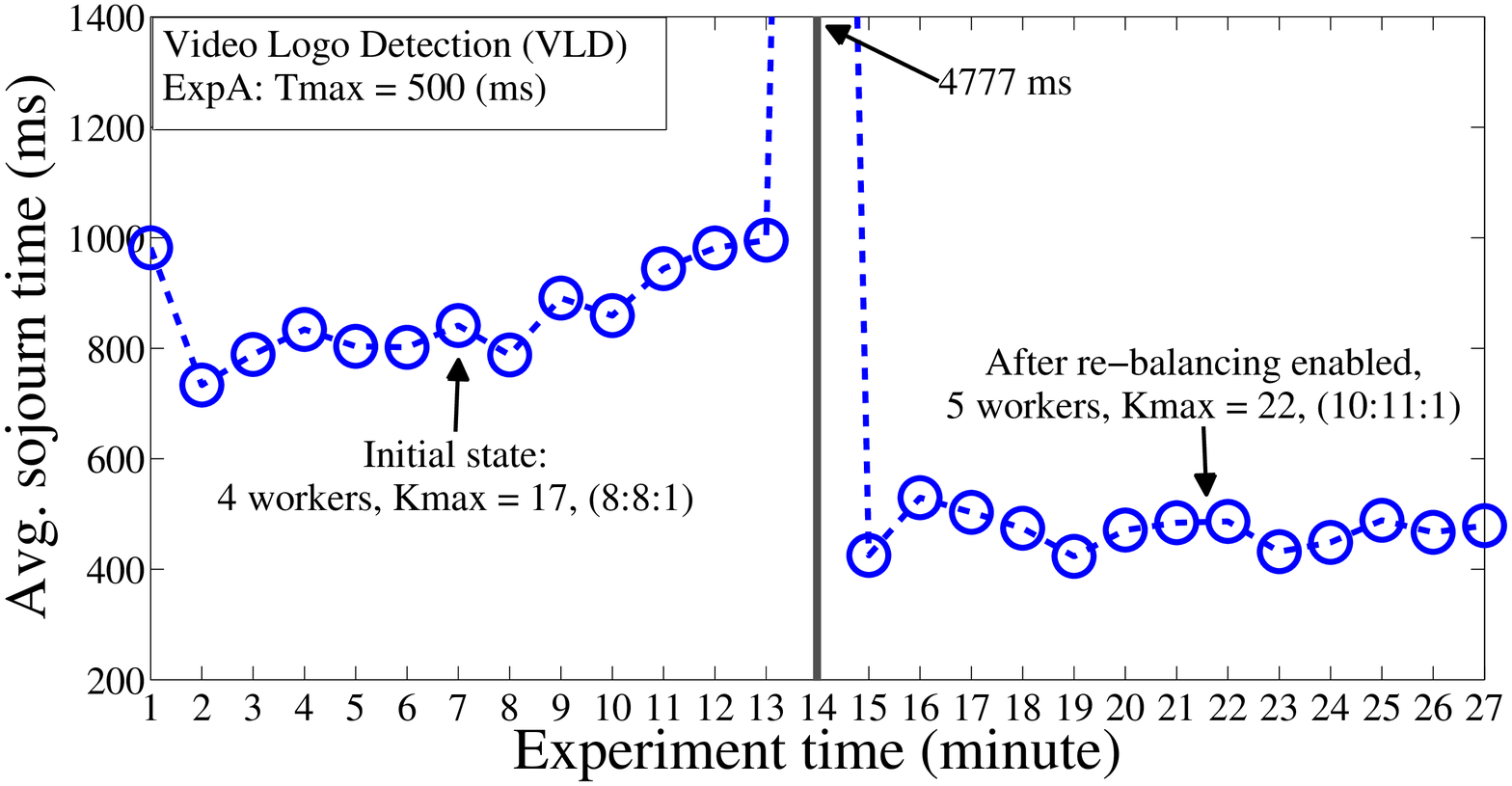}
    \includegraphics[width=2.6in]{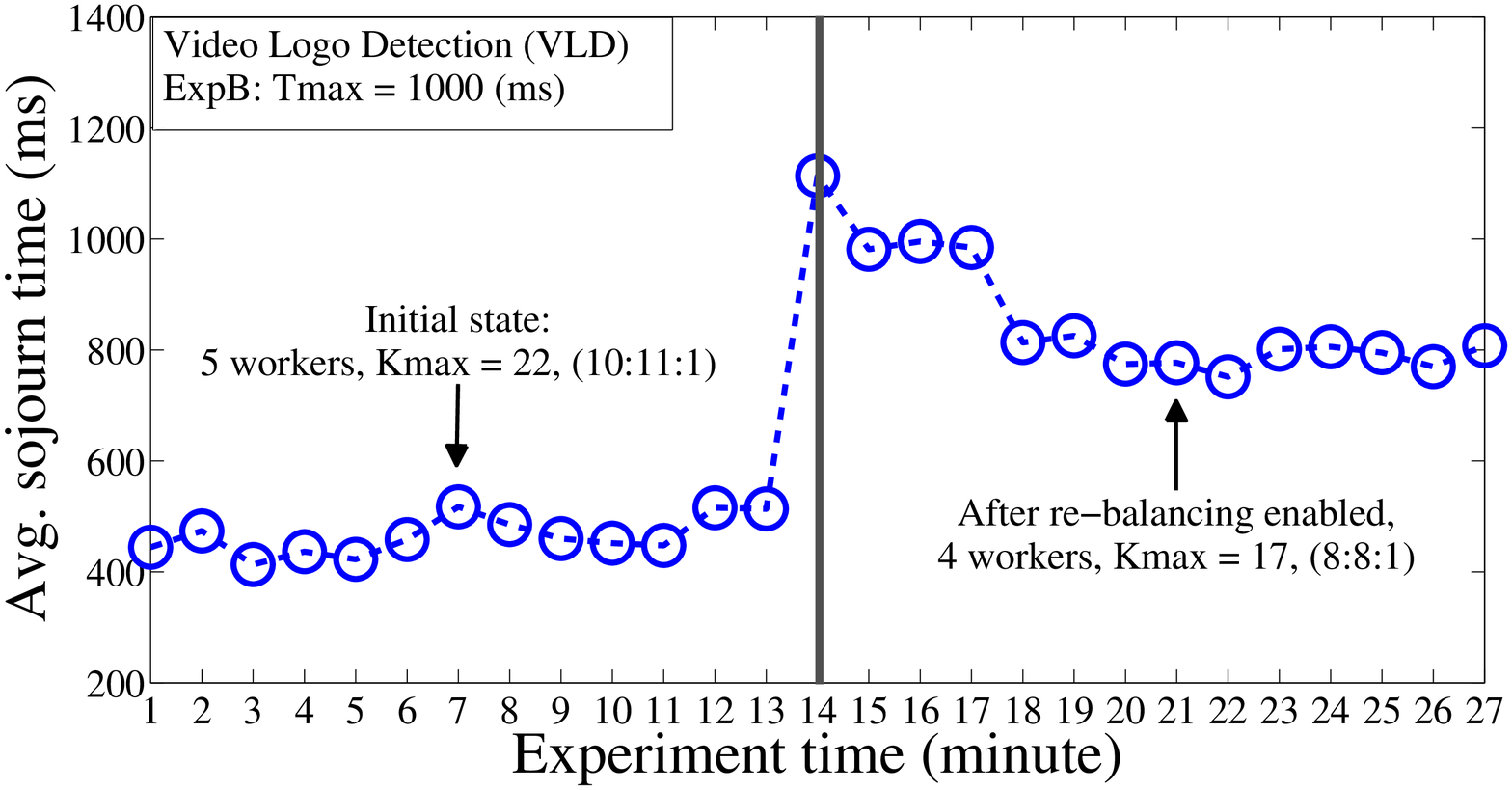}
\vspace{-5pt}
    \caption{The average sojourn times under two configurations of the VLD application, where re-balancing function is disabled from the beginning till the end of the 13th minute, and becoming and keeping enabled since the 14th minute. For ``ExpA'',    we set $T_{\max} = 500$ (ms) and in the initial state, 4 workers with $K_{\max} = 17$ executors are allocated; and for ``ExpB'', we set $T_{\max} = 1000$ (ms) and initially, 5 workers with $K_{\max} = 22$ executors are allocated.}
%\caption{The average sojourn times of three different allocations in the initial state for each application, where re-balancing function is disabled from the beginning till the end of the 13th minute, and becoming and keeping enabled since the 14th minute.}
    \label{Fig:logdetRebB}
\end{figure}

Next, we investigate how DRS adjusts resources when it detects the resource shortage/wastage according to the configured parameter $T_{\max}$. Two experiments on VLD application are conducted and each one lasts for 27 minutes and the re-balancing function is disabled from the beginning till the end of the 13th minute, and becomes enabled afterwards. The average tuple complete sojourn time of the two experiments in each minute is plotted in Figure~\ref{Fig:logdetRebB}. In particular, for ``ExpA'', we set $T_{\max} = 500$ (ms) and in the initial state, 4 workers with $K_{\max} = 17$ are allocated; and for ``ExpB'', we set $T_{\max} = 1000$ (ms) and initially, 5 workers with $K_{\max} = 22$ are allocated.

As shown in Figure~\ref{Fig:logdetRebB}, the curve of ``ExpA'' keeps with the allocation configuration (8:8:1) which is actually the suggested allocation when $K_{\max} = 17$ (by DRS algorithm), for the first 13 minutes. It has experienced the larger average tuple complete sojourn time than the configured $T_{\max} = 500$ (ms). On the 14th minute right after the re-balancing function is enabled, the DRS quickly triggers the re-scheduling operation including: (a) initializing and adding an extra machine (thus 5 more executors); and (b) calculating the recommended allocation configuration (10:11:1) when $K_{\max}$ becomes 22. Since then, the curve of ``ExpA'' is stably below the target requirement of $T_{\max} = 500$ (ms). On the other hand, the curve of ``ExpB'' shows a totally opposite shape to that of ``ExpA'', which is just as expected: it initially keeps with the configuration of (10:11:1) till the end of the 13th minutes. Afterwards, DRS triggers the re-balancing operation and makes ``ExpB'' using less resources, i.e., 4 machines, $K_{\max} = 17$ and (8:8:1), but still satisfying the performance requirement $T_{\max} = 1000$ (ms).

Similar to the observations we made on Figure~\ref{Fig:logdetMfpReb}, the cost incurred by our improved version of the re-balancing mechanism in ``ExpA'' and ``ExpB'' are again much lower than that of Storm's default version, as demonstrated by Figure~\ref{Fig:logdetRebB} . Particularly, ``ExpB'' just experiences an increase to about 1113 (ms) in average sojourn time in the 14th minute, whereas the overhead of ``ExpA'' is larger, an increase to around 4777 (ms). This is mainly because of the different actions taken during the re-scheduling, i.e., in ``ExpA'', new machines are initialized and added to the running topology, in which case, \emph{reusing} JVMs has no effects; in contrast in ``ExpB'', it only needs to stop and remove some existing working machines. Therefore, there is still room for improvement on our version of the re-balancing mechanism, which we consider as the future work.

%However, according to the four curves experiencing the re-scheduling events shown in Figure~\ref{Fig:logdetMfpReb}, (8:12:2) and (11:9:2) of VLD and (8:12:2) and (7:13:2) of FPD, the actual ``re-balancing'' operation produced surprisingly low cost, e.g., a neglectable increment in the average sojourn time within the 14th minute only. Specifically, the increased sojourn time of the curve (8:12:2) in VLD is even smaller than the original ones. Because in these experiments, we applied our own implementation (which involves coding at the Storm-core layer with Clojure) of the ``re-balancing'' mechanism, with considerable improvements against the default version. The discussion on the details of our implementation of the ``re-balancing'' function is out of the scope of this paper. The most essential improvement we have made is to re-use the JVMs by avoiding shutting down and restarting the worker processes.

\noindent\textbf{The running overhead of the DRS.}
To evaluate the computation overhead of the overall DRS layer, we report the CPU time spent by the whole DRS module, including the processing on measurement results and calculating the optimal allocation. In this experiment, we only test on the video logo detection topology composed by three bolt operators with all the parameters, $\lambda_0$, $\lambda_i$ and $\mu_i, i = 1, 2, 3$ fixed. We try different $K_{\max}$, i.e. total number of executors for all operators. For each value of $K_{\max}$, we run the procedure 100,000 times and report the average running time of the whole DRS layer. The results are listed in Table~\ref{Tab:overhead}, with \emph{Scheduling} as the allocation computation and \emph{Measurement} as the metric processing computation.
\begin{table}[htb]
\centering \caption{Computation overheads in milliseconds under different $K_{\max}$.}
\label{Tab:overhead}
\begin{tabular}{|c|c|c|c|c|c|}
\hline
    $K_{\max}$  & 12 &  24  &  48 &   96 & 192\\
\hline
    Scheduling & 0.083   & 0.158 & 0.323 & 0.665 & 1.250 \\
\hline
    Measurement & 0.100  & 0.100 & 0.100 & 0.100 & 0.100\\
\hline
\end{tabular}
\end{table}
Generally speaking, the computation done by DRS is almost neglectable, with overhead less than milliseconds in most of the cases. Moreover, the results are consistent with our intuition that the computation consumption is linear to $K_{\max}$, as analyzed over Algorithm~\ref{Alg:BoxmaT}. The time consumed on processing the measurement results is irrelevant to $K_{\max}$.
%In fact, it is affected by the total number of tasks of the topology, as we have mentioned early in this section that this number keeps immutable when the topology is continuously running.
In fact, it is affected by the total number of tasks of the topology, as we will discuss in Appendix \ref{appendix:platform} that this number keeps immutable when the topology is continuously running.
\section{Conclusion}\label{Sec:Conclusion}
This paper proposes DRS, a novel dynamic resource scheduler for real-time streaming analytics in a cloud-based DSMS. DRS overcomes several fundamental challenges, including the estimation of the required resources necessary for satisfying real-time requirements, effective and efficient resource provisioning and scheduling, and the efficient implementation of such a scheduler in a cloud-based DSMS. The performance model of DRS is based on rigorous queuing theory, and it demonstrates robust performance even when the underlying conditions of the theory are not fully satisfied. In addition, we have integrated DRS into a popular system Storm, and evaluated it by conducting extensive experiments based on real applications and datasets.

Regarding future work, we plan to investigate efficient strategies for migrating the system from the current resource configuration to the new one recommended by DRS. This step should minimize additional overhead and result latency during migration, as well as the migration duration, (e.g.,~\cite{dingOSM}). Another interesting direction for future work is to investigate the possibility of improving performance model accuracy with more sophisticated queuing theory.


%\section*{Acknowledgment}

\section*{Acknowledgment}

This work is supported by Human-Centered Cyberphysical
Systems (HCCS) Programme at Advanced Digital Sciences Center from
Singapore's A*STAR. Ding is partially supported by NSF of China under Grant 61173081, and also supported by GZSI's grant 2013Y2-00046. The authors would like to thank Prof. Hongyang Chao from Sun Yat-sen University for her suggestions to this work.

%
%\end{thebibliography}
\bibliographystyle{IEEEtran}
\bibliography{IEEEabrv,DRS}

\begin{thebibliography}{10}
\providecommand{\url}[1]{#1}
\csname url@samestyle\endcsname
\providecommand{\newblock}{\relax}
\providecommand{\bibinfo}[2]{#2}
\providecommand{\BIBentrySTDinterwordspacing}{\spaceskip=0pt\relax}
\providecommand{\BIBentryALTinterwordstretchfactor}{4}
\providecommand{\BIBentryALTinterwordspacing}{\spaceskip=\fontdimen2\font plus
\BIBentryALTinterwordstretchfactor\fontdimen3\font minus
  \fontdimen4\font\relax}
\providecommand{\BIBforeignlanguage}[2]{{%
\expandafter\ifx\csname l@#1\endcsname\relax
\typeout{** WARNING: IEEEtran.bst: No hyphenation pattern has been}%
\typeout{** loaded for the language `#1'. Using the pattern for}%
\typeout{** the default language instead.}%
\else
\language=\csname l@#1\endcsname
\fi
#2}}
\providecommand{\BIBdecl}{\relax}
\BIBdecl

\bibitem{lam2012muppet}
W.~Lam, L.~Liu, S.~Prasad, A.~Rajaraman, Z.~Vacheri, and A.~Doan, ``Muppet:
  Mapreduce-style processing of fast data,'' \emph{Proc. of the VLDB
  Endowment}, vol.~5, no.~12, pp. 1814--1825, 2012.

\bibitem{qian2013timestream}
Z.~Qian, Y.~He, C.~Su, Z.~Wu, H.~Zhu, T.~Zhang, L.~Zhou, Y.~Yu, and Z.~Zhang,
  ``Timestream: Reliable stream computation in the cloud,'' in \emph{Proc. of
  ACM European Conference on Computer Systems}, 2013.

\bibitem{dean2008mapreduce}
J.~Dean and S.~Ghemawat, ``Mapreduce: simplified data processing on large
  clusters,'' \emph{Comm. of ACM}, vol.~51, no.~1, pp. 107--113, 2008.

\bibitem{li2014distributed}
F.~Li, B.~C. Ooi, M.~T. {\"O}zsu, and S.~Wu, ``Distributed data management
  using mapreduce,'' \emph{ACM Computing Surveys}, vol.~46, no.~3, p.~31, 2014.

\bibitem{fairScheduler}
\url{http://hadoop.apache.org/docs/r1.2.1/fair_scheduler.html}.

\bibitem{capacityScheduler}
\url{http://hadoop.apache.org/docs/r1.2.1/capacity_scheduler.html}.

\bibitem{zaharia2010delay}
M.~Zaharia, D.~Borthakur, J.~Sen~Sarma, K.~Elmeleegy, S.~Shenker, and
  I.~Stoica, ``Delay scheduling: a simple technique for achieving locality and
  fairness in cluster scheduling,'' in \emph{Proc. of the European conference
  on Computer systems}, 2010, pp. 265--278.

\bibitem{melnik2010dremel}
S.~Melnik, A.~Gubarev, J.~J. Long, G.~Romer, S.~Shivakumar, M.~Tolton, and
  T.~Vassilakis, ``Dremel: interactive analysis of web-scale datasets,''
  \emph{Proc. of the VLDB Endowment}, vol.~3, no. 1-2, pp. 330--339, 2010.

\bibitem{presto}
M.~Traverso, ``Presto: Interacting with petabytes of data at facebook.''

\bibitem{zhang2014oceanrt}
S.~Zhang, Y.~Yang, W.~Fan, L.~Lan, and M.~Yuan, ``Oceanrt: Real-time analytics
  over large temporal data,'' in \emph{ACM SIGMOD, Demo}, 2014.

\bibitem{Zhang2014oceanrt2}
S.~Zhang, Y.~Yang, W.~Fan, and M.~Winslett, ``Design and implementation of a
  real-time interactive analytics system for large spatio-temporal data,''
  \emph{Proc. of the VLDB Endowment}, vol. 7(13), pp. 1754--1759, 2014.

\bibitem{zhang2013c}
Z.~Zhang, H.~Shu, Z.~Chong, H.~Lu, and Y.~Yang, ``C-cube: Elastic continuous
  clustering in the cloud,'' in \emph{Proc. of IEEE ICDE}, 2013.

\bibitem{cai2013sada}
R.~Cai, Z.~Zhang, and Z.~Hao, ``Sada: A general framework to support robust
  causation discovery,'' in \emph{Proc. of ICML}, 2013, pp. 208--216.

\bibitem{thusoo2010hive}
A.~Thusoo, J.~S. Sarma, N.~Jain, Z.~Shao, P.~Chakka, N.~Zhang, S.~Antony,
  H.~Liu, and R.~Murthy, ``Hive-a petabyte scale data warehouse using hadoop,''
  in \emph{Proc. of IEEE ICDE}, 2010, pp. 996--1005.

\bibitem{hindman2011mesos}
B.~Hindman, A.~Konwinski, M.~Zaharia, A.~Ghodsi, A.~D. Joseph, R.~H. Katz,
  S.~Shenker, and I.~Stoica, ``Mesos: A platform for fine-grained resource
  sharing in the data center.'' in \emph{Proc. of USENIX NSDI}, 2011.

\bibitem{yarn}
\url{http://hadoop.apache.org/docs/current/hadoop-yarn/hadoop-yarn-site/}.

\bibitem{zhang2013abacus}
Z.~Zhang, R.~T.~B. Ma, J.~Ding, and Y.~Yang, ``Abacus: An auction-based
  approach to cloud service differentiation,'' in \emph{Proc. of IEEE
  International Conference on Cloud Engineering}, 2013, pp. 292--301.

\bibitem{arasu2003stream}
A.~Arasu, B.~Babcock, S.~Babu, M.~Datar, K.~Ito, I.~Nishizawa, J.~Rosenstein,
  and J.~Widom, ``Stream: the stanford stream data manager (demo),'' in
  \emph{Proc. of the ACM SIGMOD}, 2003, pp. 665--665.

\bibitem{arasu2006cql}
A.~Arasu, S.~Babu, and J.~Widom, ``The cql continuous query language: semantic
  foundations and query execution,'' \emph{The International Journal on Very
  Large Data Bases}, vol.~15, no.~2, pp. 121--142, 2006.

\bibitem{babu2005adaptive}
S.~Babu, K.~Munagala, J.~Widom, and R.~Motwani, ``Adaptive caching for
  continuous queries,'' in \emph{Proc. of IEEE ICDE}, 2005, pp. 118--129.

\bibitem{abadi2003aurora}
D.~J. Abadi, D.~Carney, U.~{\c{C}}etintemel, M.~Cherniack, C.~Convey, S.~Lee,
  M.~Stonebraker, N.~Tatbul, and S.~Zdonik, ``Aurora: a new model and
  architecture for data stream management,'' \emph{The International Journal on
  Very Large Data Bases}, vol.~12, no.~2, pp. 120--139, 2003.

\bibitem{cranor2003gigascope}
C.~Cranor, T.~Johnson, O.~Spataschek, and V.~Shkapenyuk, ``Gigascope: a stream
  database for network applications,'' in \emph{Proc. of the ACM SIGMOD}, 2003,
  pp. 647--651.

\bibitem{chandrasekaran2003telegraphcq}
S.~Chandrasekaran, O.~Cooper, A.~Deshpande, M.~J. Franklin, J.~M. Hellerstein,
  W.~Hong, S.~Krishnamurthy, S.~R. Madden, F.~Reiss, and M.~A. Shah,
  ``Telegraphcq: continuous dataflow processing,'' in \emph{Proc. of the ACM
  SIGMOD}, 2003, pp. 668--668.

\bibitem{andrade2011processing}
H.~Andrade, B.~Gedik, K.-L. Wu, and P.~Yu, ``Processing high data rate streams
  in system s,'' \emph{Journal of Parallel and Distributed Computing}, vol.~71,
  no.~2, pp. 145--156, 2011.

\bibitem{babcock2004operator}
B.~Babcock, S.~Babu, M.~Datar, R.~Motwani, and D.~Thomas, ``Operator scheduling
  in data stream systems,'' \emph{The International Journal on Very Large Data
  Bases}, vol.~13, no.~4, pp. 333--353, 2004.

\bibitem{abadi2005design}
D.~J. Abadi, Y.~Ahmad, M.~Balazinska, U.~Cetintemel, M.~Cherniack, J.-H. Hwang,
  W.~Lindner, A.~Maskey, A.~Rasin, E.~Ryvkina \emph{et~al.}, ``The design of
  the borealis stream processing engine.'' in \emph{Proc. of Conference on
  Innovative Data Systems Research}, vol.~5, 2005, pp. 277--289.

\bibitem{zaharia2012discretized}
M.~Zaharia, T.~Das, H.~Li, S.~Shenker, and I.~Stoica, ``Discretized streams: an
  efficient and fault-tolerant model for stream processing on large clusters,''
  in \emph{Proc. of the USENIX conference on Hot Topics in Cloud
  Ccomputing}.\hskip 1em plus 0.5em minus 0.4em\relax USENIX Association, 2012,
  pp. 10--10.

\bibitem{storm}
A.~Toshniwal, S.~Taneja, A.~Shukla, K.~Ramasamy, J.~M. Patel, S.~Kulkarni,
  J.~Jackson, K.~Gade, M.~Fu, J.~Donham \emph{et~al.}, ``Storm@ twitter,'' in
  \emph{Proc. of ACM SIGMOD}, 2014, pp. 147--156.

\bibitem{neumeyer2010s4}
L.~Neumeyer, B.~Robbins, A.~Nair, and A.~Kesari, ``S4: Distributed stream
  computing platform,'' in \emph{Proc. of IEEE International Conference on Data
  Mining Workshops (ICDMW)}, 2010, pp. 170--177.

\bibitem{samza}
SAMZA, \url{http://samza.incubator.apache.org/}.

\bibitem{soliman2014orca}
M.~A. Soliman, L.~Antova, V.~Raghavan, A.~El-Helw, Z.~Gu, E.~Shen, G.~C.
  Caragea, C.~Garcia-Alvarado, F.~Rahman, M.~Petropoulos \emph{et~al.}, ``Orca:
  a modular query optimizer architecture for big data,'' in \emph{Proc. of the
  ACM SIGMOD}, 2014, pp. 337--348.

\bibitem{mitchell2013using}
C.~Mitchell, Y.~Geng, and J.~Li, ``Using one-sided rdma reads to build a fast,
  cpu-efficient key-value store,'' in \emph{USENIX Annual Technical
  Conference}, 2013, pp. 103--114.

\bibitem{collins2009flexible}
R.~L. Collins and L.~P. Carloni, ``Flexible filters: load balancing through
  backpressure for stream programs,'' in \emph{Proc. of the seventh ACM
  international conference on Embedded software}, 2009, pp. 205--214.

\bibitem{xing2005dynamic}
Y.~Xing, S.~Zdonik, and J.-H. Hwang, ``Dynamic load distribution in the
  borealis stream processor,'' in \emph{Proc. of IEEE ICDE}, 2005, pp.
  791--802.

\bibitem{bitran1996state}
G.~R. Bitran and R.~Morabito, ``State-of-the-art survey: Open queueing
  networks: Optimization and performance evaluation models for discrete
  manufacturing systems,'' \emph{Production and Operations Management}, vol.~5,
  no.~2, pp. 163--193, 1996.

\bibitem{erlang1917solution}
A.~K. Erlang, ``Solution of some problems in the theory of probabilities of
  significance in automatic telephone exchanges,'' \emph{Elektrotkeknikeren},
  vol.~13, pp. 5--13, 1917.

\bibitem{tijms1986stochastic}
H.~C. Tijms, \emph{Stochastic modelling and analysis: a computational
  approach}.\hskip 1em plus 0.5em minus 0.4em\relax John Wiley \& Sons, Inc.,
  1986.

\bibitem{jackson1963jobshop}
J.~R. Jackson, ``Jobshop-like queueing systems,'' \emph{Management science},
  vol.~10, no.~1, pp. 131--142, 1963.

\bibitem{boxma1990machine}
O.~Boxma, A.~Rinnooy~Kan, and M.~Van~Vliet, ``Machine allocation problems in
  manufacturing networks,'' \emph{European Journal of Operational Research},
  vol.~45, no.~1, pp. 47--54, 1990.

\bibitem{lindeberg2012scale}
T.~Lindeberg, ``Scale invariant feature transform,'' \emph{Scholarpedia},
  vol.~7, no.~5, p. 10491, 2012.

\bibitem{AS94}
D.~Burdick, M.~Calimlim, and J.~Gehrke, ``Mafia: A maximal frequent itemset
  algorithm for transactional databases,'' in \emph{Proc. of IEEE ICDE}, 2001,
  pp. 443--452.

\bibitem{dingOSM}
\BIBentryALTinterwordspacing
J.~Ding, T.~Z.~J. Fu, R.~T.~B. Ma, M.~Winslett, Y.~Yang, Z.~Zhang, and H.~Chao,
  ``Optimal operator state migration for elastic data stream processing,''
  Tech. Rep., Feb. 2015. [Online]. Available:
  \url{http://arxiv.org/abs/1501.03619}
\BIBentrySTDinterwordspacing

\end{thebibliography}

\appendices
\section{Proof of Theorem \ref{theroem1}}\label{appendix:theroemProof}
\begin{proof}
(sketch): Let $\mathbf{k}$ be the output of $AssignProcessors$ shown in Algorithm~\ref{Alg:BoxmaT}, and $\mathbf{k}^*$ be an optimal assignment that minimizes $E[T]$. Choose any two operators $x$ and $y$ satisfying that $k^*_x>k_x$ and $k^*_y<k_y$. According to the facts that (a) $AssignProcessors$ always increments the number of processors for the operators with the highest marginal benefit (lines 10 - 14); and (b) the diminishing marginal benefit property in Inequality~(\ref{EQ:ETiConvex}), we derive the following inequality:
\begin{equation}
\label{EQ:ETiConvex4XY}
\lambda_y\Big[E[T_y](k^*_y) - E[T_y](k^*_y + 1)\Big] \geq \lambda_x\Big[E[T_x](k^*_x - 1) - E[T_x](k^*_x)\Big]\nonumber
\end{equation}
In other words, in $\mathbf{k}^*$, taking one processor away from operator $x$ and assigning it to operator $y$ leads to a value of $E[T]$ that is no worse than before. This can be done repeatedly to gradually change $\mathbf{k}^*$ to $\mathbf{k}$, without increasing $E[T]$. Hence, $E[T](\mathbf{k})\leq E[T](\mathbf{k}^*)$. Since $\mathbf{k}^*$ is optimal, $\mathbf{k}$ must be optimal as well.
\end{proof}

\section{Technical details of the three DRS modules}\label{appendix:drsModules}
\subsection{Measurer Module}

The measurer module is mainly responsible for the measurement on the CSP layer and the pre-processing of the metrics before sending them to the optimizer component. Recall from Algorithm~\ref{Alg:BoxmaT} in Section \ref{sec:drs:algo} that for each operator $i$ of an application running on the CSP layer, it is essential to collect two local metrics of the operator: the average aggregate tuple arrival rate, denoted by $\hat{\lambda_i}$, and the average service rate, denoted by $\hat{\mu_i}$. In addition, the optimizer component also needs certain global metrics for its optimization algorithm, i.e., metrics related to individual tuples and multiple operators, which include the average arrival rate of external tuples, denoted by $\hat{\lambda_0}$, and the average tuple complete sojourn time, $E[\hat{T}]$, as described in Section~\ref{sec:drs:algo}.
%denoted by $E[\hat{T}]$. Note that the performance model derived in previous section allows the system to estimate $E[T]$ based on the other metric. We, however, believe it helps to reduce the error between the estimated and actual sojourn time by directly measuring on the infrastructure.

There are two major technical challenges to the measurer module in DRS layer. Firstly, the operators and the instances within the operators may run on different physical machines during online stream processing. Therefore, the measurement must be conducted collaboratively in a distributed environment. Secondly, it is important for the measurer module to minimize the overhead of the measurement itself and maintain the high availability of the streaming processing service.
%\footnote{By queueing theory, when the system is in the steady state, the system throughput is equal to the external tuple (processing tree) arrival rate, $\lambda_0$}.

To tackle the challenges listed above, the measurer module in our system is designed as an independent system operator, mostly invisible to the system user and programmer. To collect the local metrics, a group of optional measurement logics are injected into the executables on each instance of the operators, such that specified local metrics are forcefully collected and kept in the memory of the distributed nodes. A pull-based mechanism is employed to control the data flow from the operators of the topology to the measurement operator. To limit the overhead of distributed metric collection, a bi-layer sampling strategy is applied to the system. Specifically, each instance of the operators records the metric of a tuple every $N_m$ local input tuples, while the centralized measurement operator pulls updates from the other operators every $T_m$ seconds.

To collect the global metric with respect to external tuples coming into the system, the measurement operator tracks the processing tree of the tuple, using existing techniques, e.g. acknowledgment mechanism. Therefore, the measurement operator receives notifications from the underlying infrastructure on the completion of processing tree of the external tuples, and thus retrieves global metrics based on the notification time.

After the collection of original metrics, the system still needs to go through pre-processing operations to eliminate the effects of noises, message loss and outliers. The operations include:
(a) result aggregation at the operator level. This is crucial because the metrics we have defined and are interested in are at the operator level (e.g., the Jackson network), rather than the instance level, which may only contain some proportion; and
(b) results smoothing. It helps to reduce the effects of noise and improve stability of the system. There are two options of smoothness operations supported in our system. $d(n)$ is used to denote the measurement results of the $n$th interval collected and aggregated by the controller, and $D(n)$ is used to denote the smoothed results after the $n$th interval. The first smoothness option is \emph{$\alpha$-weighed averaging}, in which we have $D(n) = \alpha D(n-1) + (1-\alpha)d(n)$, with $\alpha\in[0, 1)$ as a tunable parameter controlling the fading rate of old metrics. The second smoothness option is \emph{window-based averaging}, in which we have $D(n) = \frac{1}{w}\sum_{j=n-w+1}^n d(j)$, with $w$ as the windows size parameter.

%At last, the measurer sends the measurement results after proper handling to the optimizer, either periodically (push-based) or being notified (pull-based).

\subsection{Scheduler and Negotiator Modules}

Based on the optimization model specified by the user, the measurer returns two types of optimization results, which minimize the latency based on available resource and minimize the computation resource based on the maximal latency. Since the optimization output only indicates the amount of resources assigned to the particular operators, the system could execute the results, only when a concrete mapping between the available resource and operator is constructed. The scheduler module and resource negotiator module thus play important roles to work as a translator in the system architecture.

%\paragraph{The scheduler.}
The output of Algorithm~\ref{Alg:BoxmaT} is an optimal solution, given the currently maximum available resources $K_{\max}$. The scheduler first checks if the optimal solution is the same as the current allocation, which is read through the configuration reader as a type of input parameter. If not, then the scheduler will trigger the ``resource allocation'' module in the CSP layer, to conduct a re-allocation. One technical difficulty is that the implementation of the scheduler must be coupled closely with the CSP layer, e.g., the API calls and how the ``resource allocation'' works, as well.

%\paragraph{The re-allocation cost.}
In a practical CSP system, resource allocation always incur costs, such as processing data migration, intermediate state save and load, etc. In the long run, the optimal solution of Program~(\ref{opt1}) may adapt, due to the change of the input data rate, or change of the data properties which further affect the service time of each component. It is also possible that the current allocation becomes sub-optimal, %due to the input data change,
but its performance in terms of the tuples' average complete sojourn time is not far from the best that can be achieved.

In consequence, it is necessary to make another decision by the scheduler, i.e., given the optimal allocation and its expected performance (derived through our analytical queueing model), and the currently working allocation and the measured average complete sojourn time, and considering the cost (input as a parameter), whether it is beneficial enough to make the reallocation happen.

%\paragraph{The negotiator.}

On the other hand, finding the minimum required amount of resources, i.e., Program~(\ref{opt2}) is meaningless if we interpret the output only at the logical resource unit (pool) level, but meaningful at the physical resource layer.
The motivation of doing this must be explained in a practical way, e.g., with less available physical resources (which are controlled by the resource manager of the CSP layer), it saves costs, in terms of: a) expenditure for renting the virtual machines from the cloud services especially when the budget is tight; and b) power consumption, e.g., for local machines.

Therefore, the resource negotiator works at an even lower layer than the ``resource manager'' of the CSP layer. It negotiates with the physical machines or the cloud service provider by implementing several dedicated APIs, e.g., one of the most important must be launching/stopping the ``resource manager'' daemon process of the CSP layer.

\subsection{Configuration Reader Module}

The configuration reader is designed to be a general interface for managing a data structure containing the configuration parameters provided by either the users or the CSP layer.
We list part of the parameters: a) the type of the optimization problem, i.e., Program~(\ref{opt1}) or Program~(\ref{opt2}); b) the corresponding $K_{\max}$ and $T_{\max}$ for the algorithms in the optimizer; c) for the measurer, e.g., sampling rate $N_m$, trigger interval $T_m$ and $\alpha$ or $w$ for the smooth processing; d) for the scheduler, e.g., the current running allocation of the CSP layer, and the re-allocation cost.

\section{Overview of Storm and our implementation of the DRS modules}\label{appendix:platform}
An application running on Storm is defined by a \emph{topology}, with vertices as user-defined operators (containing computation logics) and edges as indicators of data flows between operators. %Note that Storm supports general topologies with loops, as well as multiple upstream operator and downstream operator. This feature makes existing scheduling strategy difficult to work on Storm. (David: if you want to make this point you have to explain what and why existing scheduling strategies don't work)
There are two types of operators in Storm, \emph{spouts} and \emph{bolts}. A spout acts as a data source, which connects to external streaming sources. Bolts include all other (i.e., non-source) operators. Each operator contains one or more processors, called \emph{executors}, running on different servers in the cloud.

Storm supports dynamically ``re-scaling'' an operator (spout or a bolt), which changes its number of executors. This is implemented by decoupling the routing logics from the computation logics. The routing logics remain the same even when new executors are added. Storm's implementation is based on a partitioning scheme on each operator (spout or bolt), in which each partition is called a \emph{task}. When an operator scales out (respectively in), the number of executors of the operators increases (decreases), with the tasks reassigned to the executors. In particular, there are different partitioning rules supported by Storm, e.g., shuffle, field and direct grouping. We refer the reader to \cite{storm} for details of the partitioning rules.

Given the architecture of Storm system, resource allocation/re-allocation can be controlled by assigning different numbers of executors to operators. Storm also provides an internal mechanism for migrating to a new resource configuration, called \emph{re-balancing}. Simply put, the re-balancing mechanism suspends the entire system (e.g., by shutting down all the Java Virtual Machines), modifies the executor to operator mappings and routing, and finally resumes the system. Hence, the response time becomes very high during re-balancing.
Therefore, in the real implementation of DRS and the experiment, we developed and used our own version (which involves coding at the Storm core layer in Clojure) of the re-balancing mechanism, with significant improvements over Storm's default version. Discussions on how to migrate to a new resource configuration without such costly system-wide suspensions is out of the scope of this paper. The most essential improvement we have made is to re-use the JVMs.
Finally, Storm provides a scheduler interface that enables customized executor assignment strategy, and allows users to specify the operation frequency of the scheduler.

\noindent\textbf{Measurer: }We implemented two new system operators (not visible to users) into the Storm system, called \emph{MeasurableSpout} and \emph{MeasurableBolt}. They wrap a normal bolt/spout, and add measurement logics. The measurement for bolts mainly records the elapsed time volumes the execution function spends on each of the incoming tuples. These measured results are collected periodically by the ``DRSMetricCollecor'' module, which is implemented using the Measurement APIs provided by Storm. To measure the queue related metrics, e.g., the average tuple arrival rate to each operator $i$, is more complicated, because there is no available API we can make use of. Therefore, we had to modify the source code of the Storm core to add the measurement logics. Note the rate measurement position should be at the tail of the operator queue, instead of the queue head.

\noindent\textbf{Scheduler: } Since Storm provides the scheduling APIs, our testing platform simply calls these APIs, which reassigns the executors, calls our version of the re-balancing function, and continues processing the incoming data stream automatically.

\noindent\textbf{Configuration Reader: }Similarly, the configuration reader reuses the APIs of the Storm system, which shares the configuration in Zookeeper.

\noindent\textbf{Negotiator: }The negotiator is at a lower level than the resource manager of the Storm. It is in charge of starting/shutting down extra/existing physical resources (e.g., physical machines or virtual machines). Our negotiator module is based on the APIs of YARN, on top of a Hadoop cluster.

\end{document}